\documentclass[pra,aps,twocolumn,showpacs,amsmath,amssymb,floatfix]{revtex4-1}
\usepackage{epsfig}
\usepackage{amssymb,amsmath}
\usepackage{color}
\usepackage{graphicx}

\newcommand{\bhat}[3]{\hat{b}_{#1 #2 }^{#3}}

\newcommand{\bhatp}[4]{\hat{b}_{#1 #2 #3 }^{#4}}
\newcommand{\nhatp}[3]{\hat{n}_{#1 #2 #3}}
\newcommand{\xvec}{\mathrm{\bf x}}
\newcommand{\hc}{\mathrm{h.c.}}

\begin{document}

\title{Macroscopic superposition states of ultracold  bosons in a double-well potential}

\author{M.~A. Garcia-March$^{1,2}$, D. R. Dounas-Frazer$^{3}$, and Lincoln D. Carr$^{1,4}$}
\address{$^1$Department of Physics, Colorado School of Mines, Golden, CO 80401, U.S.A.}
\address{$^2$Department of Physics, University College Cork, Cork, Ireland}
\address{$^3$Department of Physics, University of California, Berkeley, California 94720, USA}
\address{$^4$Universit\"at Heidelberg, Physikalisches Institut, Philosophenweg 12, D-69120 Heidelberg, Germany}


\begin{abstract}
We present a thorough description of the physical regimes for ultracold bosons in double wells, with special attention paid to macroscopic superpositions (MSs).  We use a generalization of the Lipkin-Meshkov-Glick Hamiltonian of up to eight single particle modes to study these MSs, solving the Hamiltonian with a combination of numerical exact diagonalization and high-order perturbation theory.  The MS is between left and right potential wells; the extreme case with all atoms simultaneously located in both wells and in only two modes is the famous NOON state, but our approach encompasses much more general MSs.  Use of more single particle modes brings dimensionality into the problem, allows us to set hard limits on the use of the original two-mode LMG model commonly treated in the literature, and also introduces a new mixed Josephson-Fock regime.  Higher modes introduce angular degrees of freedom   and MS states with different angular properties.
\end{abstract}

\maketitle

\section{Introduction}
\label{sec:introduction}

The superposition of quantum states Is a postulate in quantum mechanics. When these states can be distinguished macroscopically it leads to the fundamental question of how the theory describing the physics of point particles or atoms is connected with macroscopic objects~\cite{2002LeggettJPCM}. Ultracold bosons in double wells  provide a useful realization of macroscopic superposition (MS) states, in which the distinguishable macroscopic property is the localization of atoms in  one of the wells~\cite{2007DounasFrazerPRL,2009XiaoCueCTP,2011MazzarellaPRA}. In this system, it is possible to restrict the   atoms to occupy only two single particle modes, corresponding to  their being condensed in the lowest energy state and localized spatially in one of the wells~\cite{1999DalfovoRMP,2001LeggetRMP}. In this simplified two mode picture, the main two processes are the tunneling of atoms between wells and their interaction in pairs in one of the wells. When the  tunneling energy is larger than the interaction energy, this  is an ideal system for studying  Josephson effects~\cite{1986JavanainenPRL,1997MilburnPRA,1997SmerziPRL,1998ZapataPRA,2000OstrovskayaPRA,2000Giovanazzi,2005MahmudPRA,2005AlbiezPRL,2006AnanikianPRA,2006WangPRA,2006LiPRA,2007LevyNature,2010JuliaDiazPRA}. Conversely, when   interactions dominate over   tunneling, MSs can be obtained~\cite{1998SteelPRA,2002KalosakasPRA,2007DounasFrazerPRL}.  When all $N$ atoms occupy simultaneously the two single particle states localized in each well, these MSs are known as NOON states. The tunneling dominated regime is known as the \textit{Josephson regime}, while the interaction regime is known as the  \textit{Fock regime}. In the latter, the two mode approach can cease to be sufficient to describe the system,  due to large interactions populating single particle excited states. Here, we offer a detailed study of the eigenvector spectra  in one dimensional (1D) and three dimensional (3D) double wells, with  special attention to the appearance of MS states which occur in the Fock regime, and we include more modes in our approach, as  appropriate for this regime.

\begin{figure}[h]
\includegraphics[width=6.cm]{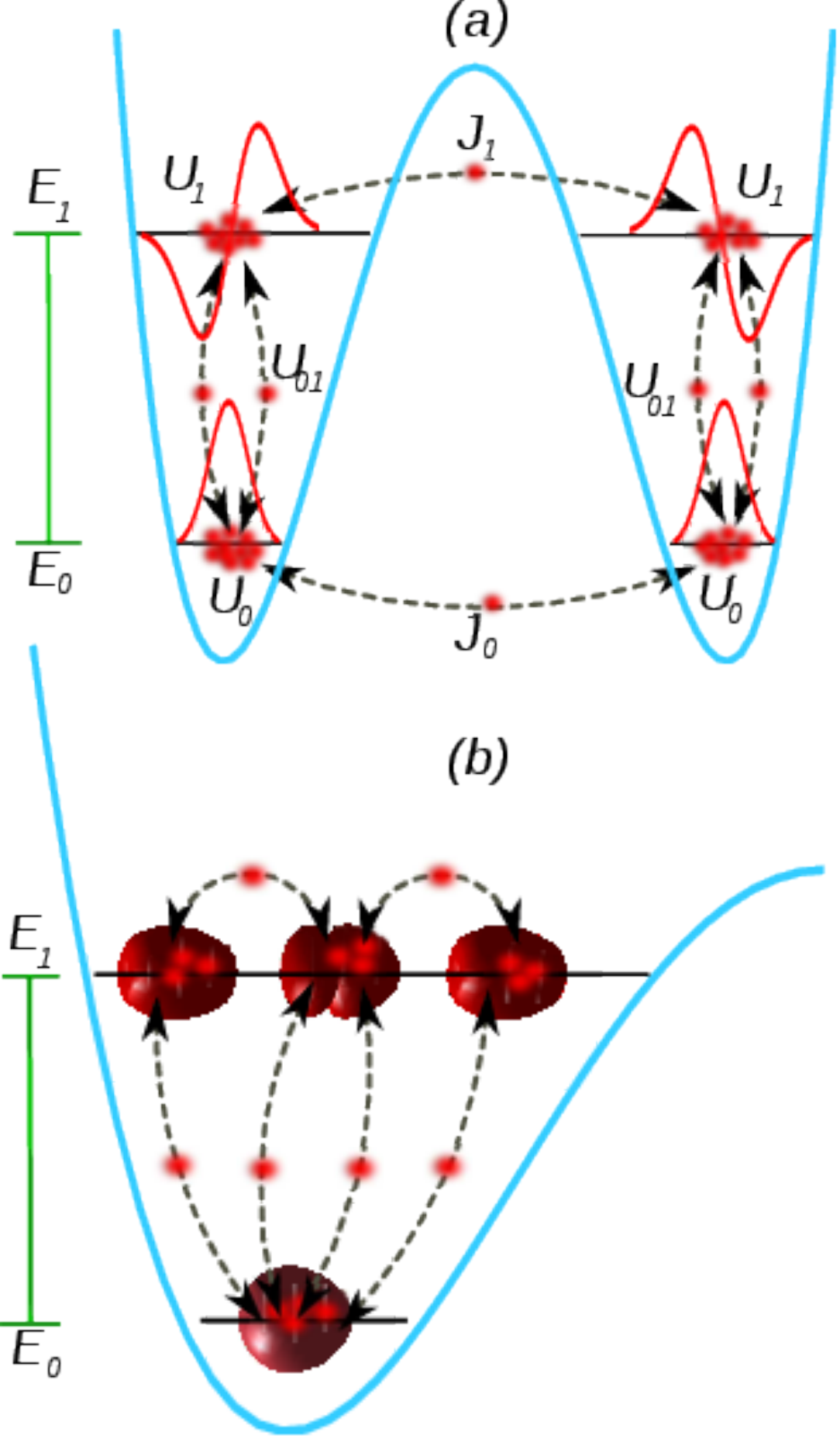}
 \vspace{-0.4cm}
 \caption{(Color online)  {\it Schematic of the Double Well.}  (a) 1D: the $J$'s indicate single-particle tunneling processes while the $U$'s indicate two-particle tunneling and interaction processes.  (b) 3D: the spherical harmonics $Y_{\ell m}$, with total and $z$-component of the angular momentum $\ell$ and $m$ respectively, are distorted by the coupling to the nearby well. \label{fig0}} \end{figure}

To model  1D double wells we use a four-mode generalization of the Lipkin-Meshkov-Glick Hamiltonian (LMGH)~\cite{1965LipkinJNP,2004VidalPRA}, where the atoms are allowed to occupy two excited on-well  (sometimes called on-site) localized single particle  eigenstates (see Fig.~\ref{fig0}(a)). Due to the new two excited modes, there are three new processes in the model. On the one hand, pairs of atoms in the ground state in one well can be  excited to the first level, or decay from this level to the ground state, as sketched in Fig.~\ref{fig0}(a). Also, the atoms excited to the first level of eigenenergies can interact on-well  or tunnel to the other well, also sketched in Fig.~\ref{fig0}(a). Then, in this picture, the conventional division  into the Josephson and Fock regimes must be amended to include a new \textit{mixed regime,} in which the atoms behave differently depending on their energy level, as discussed in Sec.~\ref{sec:bounds}. We use perturbation theory to characterize the eigenstates and eigenvalues in the Fock regime and we also characterize them analytically in the non-interacting regime. In between  these two regimes, we use  numerical exact diagonalization of the LMGH to show how the transition between these regimes occurs.  We identify clearly the regimes in which the original NOON states as well as novel MS states can be obtained.  We characterize the limits of the model, the occurrence of crossings in the spectra, and couplings between states with atoms in different levels.   Our crossing criteria clarify when the \textit{one-level approximation} (LMGH with only two modes) ceases to be sufficient, and one must use at least the \textit{two-level approximation.}

 The four-mode, two-level generalization of the LMGH is the natural way to introduce dimensionality in the problem.  When  the same two levels are generalized  to  3D double wells, an eight-mode  generalization of the LMGH is obtained as sketched in Fig.~\ref{fig0}(b). Then, a new quantum number, the orbital degree of freedom, is introduced and it plays a key role~\cite{2011GarciaMarchPRA}. Since new parameters arise in the 3D case,  new dynamical regimes are found. Moreover, new MS states involving angular degrees of freedom can occur. In a previous work we characterized the limits of the one- and two- level  approximations, the occurrence of crossings in different regimes, and the  coupling between states with atoms in different levels~\cite{2011GarciaMarchPRA}. In this Article we review, complete, and extend the 1D picture obtained in~\cite{2007DounasFrazerPRL,2007DounasFrazerMT}. We also compare with the 3D case, and summarize the derivation of the 3D two-level  LMGH and the routes to new physics in 3D which are currently being explored.

By means of a  semiclassical, mean-field approximation  macroscopic quantum tunneling   has been predicted in the Josephson regime, while macroscopic self-trapping in one of the wells  has been shown to occur when the interactions grow~\cite{1986JavanainenPRL,1997MilburnPRA,1997SmerziPRL,1998ZapataPRA,2000OstrovskayaPRA,2006AnanikianPRA,2006WangPRA,2006LiPRA,2005MahmudPRA,2010JuliaDiazPRA}.  The macroscopic quantum tunneling and self-trapping were observed in a recent experiment~\cite{2005AlbiezPRL}.  The former  can be understood as the  d.c. Josephson effect for superconductors, while the a.c. Josephson effect corresponds to   small oscillations in the difference between the populations in both wells obtained for bigger interactions~\cite{2000Giovanazzi}.   The a.c. Josephson effect was recently  demonstrated experimentally~\cite{2007LevyNature}.   One perspective is that the energy levels depend on the occupation of one of the wells. Then one can analyze the double well problem in terms of Landau-Zener tunneling. In this case the energy between levels changes with a constant small rate, leading to  an adiabatically driven Josephson junction~\cite{2000PRAWu,2002PRALiu,2006WuPRL}.

In the Fock regime the semiclassical approach  ceases to be appropriate, and other approaches can be useful; the most numerically accurate but also the most computationally intensive is the multiconfigurational Hartree method~\cite{2005MasielloPRA,2006StreltsovPRA,2008AlonPRA,2008ZollnerPRA,2009SakmanPRL}. Methods based on the LMGH~\cite{1965LipkinJNP,2004VidalPRA},  although limited to weaker interactions, have the advantage that one can treat more particles in higher dimensions, as well as make use of perturbation theory to obtain analytical results~\cite{2007DounasFrazerPRL,2010CarrEPL}. In this  Article we use  a modified LMGH for these reasons. The LMGH was first introduced in the framework of nuclear physics~\cite{1965LipkinJNP}. It has applications in a wide range of other fields, like ultracold bosons, and it is a perfect simple Hamiltonian to study a vast range of quantum effects,  from quantum phase transitions~\cite{2004VidalPRA,2005DusuelPRC}  to Josephson oscillations.  The LMGH also applies in cold atoms for two hyperfine boson species in a single well, where MS states are   predicted to occur~\cite{1998CiracPRA,2000DalvitPRA}.  In this Fock regime the two mode approach could be insufficient to characterize the problem, since crossings and coupling to other more excited single particle states may be possible. Indeed, it is necessary to consider a second level in  1D for the physical parameters found in typical BEC experimental systems~\cite{2007DounasFrazerPRL}.   To consider this possibility we use a four-mode generalization of the LMGH for 1D double wells and an eight mode characterization in the 3D case.

Besides being a good system for fundamental studies  of quantum many body effects,  ultracold bosons in double wells also have many technological applications, for example, in quantum high precision  measurements of inertial and  gravitational fields~\cite{2005SchumNatPhys,2007HallPRL},  as a primary thermometer~\cite{2006GatiNJP}, or in quantum computing~\cite{2004CalarcoPRA,2006SebbyStrableyPRA,2008StrauchPRA}.

The  Article is organized as follows. In Sec.~\ref{sec:secondQuantizedHamiltonian} we introduce the initial second quantized Hamiltonian. In Sec.~\ref{sec:1DLMG} we obtain the 1D four-mode LMGH. Then, in Sec.~\ref{sec:characterization} we characterize the eigenvectors and eigenvalues in 1D for the  Josephson and Fock regimes. In  Sec.~\ref{sec:bounds} we obtain  the limits of the model and the occurrence of crossings, we discuss all possible regimes, and we offer numerical examples to illustrate the results. Finally, we derive the 3D Hamiltonian and compare with the 1D case in  Sec.~\ref{sec:3D}. We conclude, summarize, and discuss lines of future research in Sec.~\ref{sec:conclusion}.

\section{Second Quantized Hamiltonian}
\label{sec:secondQuantizedHamiltonian}

The second quantized Hamiltonian for $N$ interacting bosons of mass $\mathcal{M}$ confined by  an external potential $V(\mathbf{x})$ with
$\mathbf{x}\in{\mathbb R}^{3}$ in terms of the  bosonic creation and annihilation field operators $\hat{\Psi}(\mathbf{x})$, $\hat{\Psi}^{\dagger}(\mathbf{x})$, obeying the usual commutation relations, is
\begin{align}
&\hat{H}  =  \int\!\!
d\mathbf{x}\,\hat{\Psi}^{\dagger}(\mathbf{x})\left[-\frac{\hbar^{2}}{2\mathcal{M}}\nabla^{2}+V(\mathbf{x})\right]\hat{\Psi}(\mathbf{x})\label{eq:second-quantized1}\\
 & +  \frac{1}{2}\int\!\! d\mathbf{x}\,\hat{\Psi}^{\dagger}(\mathbf{x})\left[\int
 d\mathbf{x'}\hat{\Psi}^{\dagger}(\mathbf{x}')V_{\mathrm{int}}(\mathbf{x}-\mathbf{x}')\hat{\Psi}(\mathbf{x}')\right]\hat{\Psi}(\mathbf{x})\,,\nonumber
\end{align}
  where $V_{\mathrm{int}}(\mathbf{x}-\mathbf{x}')$ stands for the two-body interaction and  $V(\mathbf{x})$ is a three dimensional double-well potential with
minima at $\mathbf{x}=\pm\mathbf{a}\in\mathbb{R}^{3}$ and a local maximum at $\mathbf{x}=\mathrm{{\bf 0}}$. Without loss of generality, we assume
$V(\mathbf{x})=V(x)+V(y)+V(z)$.   We take the 1D potentials in the $x$  and  $y$ directions as isotropic harmonic
oscillators, that is, $V_{x}=\frac{1}{2} \omega_{x}^2 x^2 $ and $V_{y}=\frac{1}{2} \omega_{y}^2 y^2 $, where both frequencies are of the same order. For simplicity, we consider $\omega_x=\omega_y=\omega_{\perp}$ in the following.  A conventional 1D double well potential $ V(z)$ of barrier height $V_0$  can be
approximated near its minima at $z=\pm a$
  by  \begin{equation}
V(z\pm a)\approx\frac{1}{2}\mathcal{M}\omega^{2}z^{2}\,,\label{eq:V(z)}\end{equation}
 where
 \begin{equation}
 \omega\equiv\left(\frac{1}{\mathcal{M}}\frac{\partial^2 V}{\partial z^2}\right)^{1/2}_{z=a}
  \end{equation}
is the local trapping frequency in each well. The recoil energy, defined as $E_r=\hbar^2/\mathcal{M} a^2$, is used through the dimensionless
parameter $V_0/E_r$ to determine the barrier height in most of the experiments,  with $\lambda=2a$  an effective wavelength.

We assume first that  $\omega$ is of the same order as $\omega_{\perp}$ and we consider low densities,  for which only binary collisions are relevant; also, we consider low energies, when these collisions are characterized by the
\textit{s}-wave scattering length of the atoms,  $a_s$. The diluteness condition for a weakly interacting Bose gas is \begin{equation}
\sqrt{|\bar{n}\, a_{s}^{3}|}\ll 1 \,,\end{equation}
 where $\bar{n}$ is the average density of the gas. In the context
of the double-well potential, the maximum density of the gas is approximately $\bar{n}=N/(\sqrt{2\pi}\, a_{\mathrm{ho}})^{3}$, where  $a_{\mathrm{ho}}$ is the harmonic oscillator length given by $a_{\mathrm{ho}}=\sqrt{\hbar/\mathcal{M}\omega}$ (a similar condition  must hold in the other two directions). Correspondingly, we
restrict our discussion to the regime \begin{equation} N^{1/3}\ll\left|\sqrt{2\pi}\, a_{\mathrm{ho}}/a_{s}\right|\,.\label{eq:diluteness}\end{equation}
 Although the system is said to be  weakly interacting when
condition (\ref{eq:diluteness}) is met, the interaction energy can be on the order of the kinetic energy and dilute gases can therefore exhibit
non-ideal behavior~\cite{1999DalfovoRMP,2001LeggetRMP}.  Under these conditions the second quantized Hamiltonian can be approximated as
\begin{align}
  \hat{H} =& \int\! d^3\xvec \, \hat{\Psi}^{\dagger}(\xvec)\left[ -\frac{\hbar^2}{2\mathcal{M}}\nabla^2 + V(\xvec)\right]\hat{\Psi}(\xvec)\,\nonumber\\
  & + \frac{g}{2}\int \! d^3\xvec \, \hat{\Psi}^{\dagger}(\xvec)\hat{\Psi}^{\dagger}(\xvec)\hat{\Psi}(\xvec)\hat{\Psi}(\xvec)\,.
\label{eq:second-quantized}
\end{align}
 The coupling constant $g$ is proportional to the $s$-wave scattering
length, $ g = 4\pi\hbar^2 a_s/\mathcal{M}$; in Eq.~(\ref{eq:second-quantized}) we took the two-body potential from Eq.~(\ref{eq:second-quantized1}) to be approximated by an effective local interaction $V_{\mathrm{int}}(\mathbf{x}-\mathbf{x}')=g\,\delta^{(3)}( \mathbf{x}-\mathbf{x}')$.

\section{One-dimensional Lipkin-Meshkov-Glick Hamiltonian}
\label{sec:1DLMG}

Let us consider first the reduced 1D case, taking $\omega\ll\omega_{\perp}$. The double-well potential can be reduced to one spatial  dimension in extremely anisotropic traps, where the transverse trapping frequencies must be sufficiently high to reduce the dimensionality of the single-particle wavefunctions.  However, one must avoid potential  resonances by not squeezing the trap too tightly~\cite{1998OlshaniiPRL}: it is sufficient that the transverse harmonic oscillator length  $a_{\mathrm{ho},\perp} \equiv \sqrt{\hbar/\mathcal{M}\omega_{\perp}}\gg a_s$. Under these assumptions, we can reduce the dimensionality of the second quantized Hamiltonian, by considering that the particles interact and tunnel only in one-dimension. Then, we can use appropriate superpositions of the  eigenfunctions of the one-dimensional single particle Hamiltonian
\begin{equation}
 H_{\mathrm{sp}}=-\frac{\hbar^{2}}{2\mathcal{M}}\nabla^{2}+V(z)\,,\label{eq:SPH1}
\end{equation}
to obtain  a set of on-well localized wavefunctions,
in an analogous way that  on-site Wannier states are obtained from Bloch functions on a lattice~\cite{1976Ashcroft}. We use this set of functions to expand the field operator:
 \begin{equation}
\hat{\Psi}(z)=\sum_{j,l}\hat{b}_{j\ell}\psi_{\ell}(z-z_{j})\,,\label{eq:hatPsi}
\end{equation}
 where $\psi_{\ell}(z-z_{j})$ are the on-well localized functions. Here, $z_{L}\equiv-a$ and $z_{R}\equiv a$ are the minima of the left and right wells, respectively, $j\in\{L,R\}$ is the  \textit{well  index}, and the
label $\ell$ is the  \textit{level index}. The level index increases with single particle energies in each well.  For the two level approach considered here, $\ell \in \{0,1\}$. The operators $\hat{b}_{j\ell}$ and $\hat{b}_{j\ell}^{\dagger}$ satisfy the usual bosonic annihilation and creation commutation relations,
\begin{align}
 [\hat{b}_{j\ell}^{\dagger} ,\hat{b}_{j'\ell'}] & =\delta_{jj'}\delta_{\ell\ell'}\,,\nonumber \\
{}[\hat{b}_{j\ell},\hat{b}_{j'\ell'}] & =[\hat{b}_{j\ell}^{\dagger},\hat{b}_{j'\ell'}^{\dagger}]=0\,.
\end{align}
Let  $\varphi^{n}(z)$ be the  $n$th eigenfunction of Hamiltonian~(\ref{eq:SPH1}),  with eigenvalue $\epsilon_{n}$. Then, the localized
functions at well $j$ are
\begin{align}
& \psi_{j0}(z)=\frac{1}{\sqrt{2}}\left(\varphi^{1}(z)\pm\varphi^{2}(z)\right)\,,\label{eq:sup1}\\ &
\psi_{j1}(z)=\frac{1}{\sqrt{2}}\left(\varphi^{3}(z)\pm\varphi^{4}(z)\right)\,,\label{eq:sup2}
\end{align}
with $j=L$ for the plus sign and $j=R$  for the minus sign.  The corresponding eigenvalues are  $E_{0}=(\epsilon^{1}+\epsilon^{2})/2$ and $E_{1}=(\epsilon^{3}+\epsilon^{4})/2$. These on-well localized eigenfunctions are represented schematically in Fig.~\ref{fig0}(a).

Substituting Eq.~(\ref{eq:hatPsi}) into the second quantized Hamiltonian (\ref{eq:second-quantized}) yields the two-level Hamiltonian:
\begin{equation}
\hat{H}=\hat{H}_{0}+\hat{H}_{1}+\hat{H}_{01}\label{eq:two-level}
\end{equation}
with
\begin{equation}
\label{Eq:onelevelH}
\hat{H}_{\ell}=-J_{\ell}\sum_{j\ne
j'}\hat{b}_{j\ell}^{\dagger}\hat{b}_{j'\ell}+U_{\ell\ell}\sum_{j}\hat{n}_{j\ell}\left(\hat{n}_{j\ell}-1\right) +E_{\ell}\sum_{j}\hat{n}_{j\ell}\,,\end{equation}
\begin{equation} \hat{H}_{01}=U_{01}\sum_{j,\ell\ne
\ell'}\left(2\hat{n}_{j\ell}\hat{n}_{j\ell'}+\hat{b}_{j\ell}^{\dagger}\hat{b}_{j\ell}^{\dagger}\hat{b}_{j\ell'}\hat{b}_{j\ell'}\right)\,,\end{equation}
where  $\hat{n}_{j\ell} \equiv \hat{b}_{j\ell}^{\dagger}\hat{b}_{j\ell}$ is the number operator. The hopping and interaction
terms are
\begin{equation} J_{\ell}=-\int dz \psi_{\ell}^{\ast}(z-a)\left[-\frac{\hbar^{2}}{2\mathcal{M}}\!\nabla^{2}+V(z)\right]\psi_{\ell}(z+\! a)\,
,\label{eq:J_l}
\end{equation}
 and
 \begin{equation}
U_{\ell\ell'}=\frac{g_{1}}{2}\int dz|\psi_{\ell}(z)|^{2}|\psi_{\ell'}(z)|^{2}\, ,\label{eq:U_l_l'}
\end{equation}
respectively. The coupling constant obtained after reducing the dimensionality of the problem is $g_{1}=2\hbar\omega_{\perp}a_{s}$. In the following, we denote $U_{\ell\ell}$ simply as $U_{\ell}$.

The functions $\psi_{\ell}(x)$ resemble roughly the eigenfunctions of the harmonic
oscillator potential:
 \begin{align}
\psi_{0}(z)&=a_{\mathrm{ho}}^{1/2}\pi^{-1/4}e^{-z^{2}/2a_{\mathrm{ho}}^2}\,,\label{eq:EF0}\\
\psi_{1}(z)&=a_{\mathrm{ho}}^{1/2}\sqrt{2}\,\pi^{-1/4}(z/a_{\mathrm{ho}})e^{-z^{2}/2a_{\mathrm{ho}}^2}\,.\label{eq:EF1}
\end{align}
The energy of an atom
associated to the wavefunction $\psi_{\ell}(z-z_{j})$ is
 $E_{l}\approx\hbar\omega(\ell+1/2)$.
This approximation fails when the overlap between different on-well localized eigenfunctions is small. Then, in general, it gives a poor approximation of the $J_{\ell}$ coefficients given in~(\ref{eq:J_l}). Nevertheless, they can be used to obtain a good approximation for the overlap integrals purely on-well, like those defining the $U_{\ell\ell'}$ coefficients, Eq.~(\ref{eq:U_l_l'}). This is valid for both regimes, though it gives only a rough idea of these coefficients when the actual single particle eigenfunctions are highly distorted, as  occurs in the Josephson regime. In this case, it can only be used to  obtain scaling relations in the problem. By solving these integrals analytically using Eq.~(\ref{eq:EF1}) to approximate the single particle eigenfunctions, we obtain
\begin{equation}
 \label{eq:U0}
U_{0}=\hbar\omega\frac{g_1}{\sqrt{2\pi}}\left(\frac{a_{\mathrm{ho}}}{a}\right)\,. \end{equation}
Moreover, we can relate the other interaction coefficients to $U_0$ by solving the corresponding integrals, which gives $U_{1}=(1/2)U_{0}$  and $U_{01}=(3/4)U_{0}$. Here we use these expressions in the illustrative examples presented in Sec.~\ref{sec:bounds}, though for actual potentials to obtain accurate values for these coefficients, it is necessary to calculate the single particle eigenfunctions numerically to solve the integrals~(\ref{eq:U_l_l'}).

Finally, to deduce
the Hamiltonian~(\ref{eq:two-level}) we neglected off-site interactions.
These correspond to interaction terms like:
 \begin{equation} U_{\ell \ell'}^{j
j'}=\frac{g_{1}}{2}\int dz |\psi_{\ell}(z-z_j)|^{2}|\psi_{\ell'}(z-z_{j'})|^{2}\,,\label{eq:U_llp}\end{equation}
where $j\ne j'$.  If the barrier is  infinitely high these integrals vanish. If the single particle eigenfunctions are approximated by  Eq.~(\ref{eq:EF1}) this integral gives
$U_{0}\exp[-(a/a_{\mathrm{ho}})^2]$. Then, one could assume that these terms can be neglected for sufficiently high or separated wells, that is, if  $ a_{\mathrm{ho}}\ll a$. Nevertheless, these are not on-well integrals and then the expressions of the harmonic oscillator eigenfunctions,  Eq.~(\ref{eq:EF1}) do not give accurate results. Then these integrals have to be evaluated numerically for the actual potential and for the correct numerically evaluated single particle eigenfunctions.
Moreover, these coefficients have to be compared with   $J_{\ell}$ to justify  their being neglected, since both quantities could be comparable in the Fock regime. In the Josephson regime we can safely neglect  these coefficients since they are much smaller than any other interaction coefficient, which  in turn are smaller than the tunneling ones.  For the Fock regime,  we assume that the tunneling coefficients are always bigger that these cross-terms,  and have verified that our assumption holds for a number of specific cases.  We refer the reader to Ref.~\cite{1999SpekkensPRA} for a thorough discussion  of the effect of these terms.

\section{Characterization of the Eigenstates}
\label{sec:characterization}

\subsection{General Considerations}

An arbitrary state vector $|\Psi\rangle$,  in the solution space of the two-level Hamiltonian Eq.~(\ref{eq:two-level}), can be expressed in   Fock space as
\begin{equation}
|\Psi\rangle=\sum_{i=0}^{\Omega-1}c_{i}|i\rangle,\quad|i\rangle=\bigotimes_{j,\ell}|n_{j\ell}^{(i)}\rangle\,,\label{eq:PsiFock}
\end{equation}
 where
 \begin{equation}
|n_{j\ell}^{(i)}\rangle=\frac{1}{\sqrt{n_{j\ell}!}}\left(\hat{b}_{j\ell}^{\dagger}\right)^{n_{j\ell}^{(i)}}|0\rangle\,.
\end{equation}
 Here $\Omega$ is the dimension of the Hilbert space $\{|i\rangle\}$,
$i$ is the  \textit{Fock index}, and $|c_{i}|^{2}$ is the probability of finding the atoms distributed between different energy levels and wells   in the Fock vector $|i\rangle$,
when the system is described by state $|\Psi\rangle$, that is  $c_i=\langle i|\Psi\rangle$;  henceforth we refer to $c_i$ as the \textit{Fock-space amplitude}.  The Fock vector $|i\rangle$  can also be written in longer form as
\begin{equation}
|i\rangle = |n_{L0}^{(i)},n_{R0}^{(i)}\rangle|n_{L1}^{(i)},n_{R1}^{(i)}\rangle\,,
\end{equation}
indicating the occupation of each well (L,R) and each level (0,1) associated with Fock index $i$.  We work in the canonical ensemble, i.e., we require the total number of particles
\begin{equation}
 N=\sum_{j\ell}n_{j\ell}\,,\label{eq:Ntotal}
\end{equation}
to be constant.

The  Fock index $i$ is chosen to increase  starting with the number
of atoms in well $j=L$, and then subsequently increasing with the number of atoms in the same, left well moving up into the excited level $\ell=1$. Therefore, for the first $N+1$ Fock vectors
 $i=1+n_{L0}$ and they correspond with vectors with no occupation of the excited level. Then, they satisfy
\begin{equation}
|i\rangle=\frac{1}{\sqrt{n_{L0}!n_{R0}!}}\!\left(\hat{b}_{L0}^{\dagger}\right)^{n_{L0}}
\!\left(\hat{b}_{R0}^{\dagger}\right)^{n_{R0}}|0\rangle\,,
\end{equation}
for $i=0,\;1,\;\cdots,\; N$. For example, for $N=2$, the Fock vectors with index $i=1,2,3$ are $|0,2\rangle|0,0\rangle$, $|1,1\rangle|0,0\rangle$, and  $|2,0\rangle|0,0\rangle$, respectively.
The one-level approximation can easily
be recovered from Eq.~(\ref{eq:two-level}) by requiring $i\leq N+1$. In this truncated space, the dimension of the Hilbert space reduces to that
of the one-level approximation, namely, $N+1$, and the two-level Hamiltonian $\hat{H}$ reduces to the one-level Hamiltonian $\hat{H}_{0}$.

A general expression for the Fock index is:
\begin{align} i & =  1+n_{L0}+\sum_{p=-1}^{N_{1}-1}(N+1-p)(1+p)\nonumber \\ & +(N+1-N_{1})n_{L1}\,,
\label{Eq:Fockindexordering}\end{align}
where $N_{1}$ is the number of atoms in the excited level. Then, the following Fock vectors for $N=2$ are  $|0,1\rangle|0,1\rangle$, $|1,0\rangle|0,1\rangle$, $|0,1\rangle|1,0\rangle$, and  $|1,0\rangle|1,0\rangle$ with indices $i=4,5,6,7$. The last three show occupation only of the excited level and they are $|0,0\rangle|0,2\rangle$, $|0,0\rangle|1,1\rangle$, and  $|0,0\rangle|2,0\rangle$, for $i = 8, 9, 10$.   Thus for just two atoms there are already 10 states in our two-level problem, and with increasing $N$ it is necessary to use numerical matrix methods to keep track.

The eigenstates $|\phi^{(k)}\rangle$ of the two-level Hamiltonian (\ref{eq:two-level}) satisfy:
\begin{equation}
\hat{H}|\phi^{(k)}\rangle=\varepsilon^{(k)}|\phi^{(k)}\rangle\,,\end{equation}
where $\varepsilon^{(k)}$ is the energy eigenvalue corresponding to the state
$|\phi^{(k)}\rangle$. The eigenstate label $k$ is chosen to increase with $\varepsilon^{(k)}$. In order to describe these states, we will use the previously introduced
Fock-space amplitudes \begin{equation} c_{i}^{(k)}=\left\langle i\right.|\phi^{(k)}\rangle\,,\end{equation}
now containing two separate indices for clarity: $i$ is the Fock index, describing the ordering of the Fock basis; $k$ is the energy index, describing the ordering of the energy eigenvalues.

 When interlevel effects are not relevant, the eigenstates fall into one of two categories: \emph{harmonic-oscillator-like states}, and \textit{macroscopic superposition states} (MS states). When the barrier between wells is low, $J_{0}\gg NU_{0}$, all states are harmonic oscillator-like (HO) states.  This is the Josephson regime and it is characterized by
\begin{equation}
 \xi_{J_{\ell}}=N/\zeta_{\ell}\ll 1\,,\label{eq:CA}
\end{equation}
 where $
\zeta_{\ell}=J_{\ell}/U_{\ell}$.  This criterion must be evaluated separately for both levels, since $J_1>J_0$, while the interaction terms
are of the same order.

On the other hand, MS states dominate the spectrum in the high barrier limit.   This is the Fock regime, in which
 \begin{equation}
\xi_{U_{\ell}}=\zeta_{\ell}\ll 1\,.\label{eq:CB}
\end{equation}
Again, this criterion  must be evaluated separately for the two levels.

When the level spacing is comparable to $NU_{0}$, interlevel effects  can no longer be neglected, and a third category of eigenstates emerges. These
states show weak coupling between states with particles only in the lowest energy level and others with particles in the excited one. We name these
coupled excited states \emph{shadows of the MS states}, because in a surface  plot of the Fock-space amplitudes as a function of both energy and Fock index, they appear as faint copies of the MS states with occupation  only of the lower level  at higher Fock  index~\cite{2011GarciaMarchPRA}. In the following two subsections we characterize the eigenstates in the Josephson and Fock regimes.

\subsection{Non-interacting Regime}
\label{subsec:HO}

Let us consider the extreme Josephson regime, or noninteracting regime, for which $U_{\ell}=0$. The Hamiltonian reduces to $\hat{H}=\hat{H}_{0}+\hat{H}_{1}$, with
 \begin{equation}
\hat{H}_{\ell}=-J_{\ell}\sum_{j\ne j'}\hat{b}_{j}^{\ell\dagger}\hat{b}_{j'}^{\ell}+E_{\ell}\sum_j\hat{n}_{j\ell}\,.
\label{eq:HOHam}
\end{equation}
Since the Hamiltonian is clearly separable, its
eigenstates $|\phi^{(k)}\rangle$ are a direct product of the one-level ones, $|\phi^{(K_{\ell}^{(k)})}\rangle$: \begin{equation}\label{eq:HO_states}
  |\phi^{(k)}\rangle = \bigotimes_{\ell}|\phi^{(K_{\ell}^{(k)})}\rangle\,.
\end{equation} where $K_{\ell}^{(k)}=0,1,\dots,N_{\ell}^{(k)}$ is the one-level eigenstate label and for the $k$th eigenstate there are $N_{\ell}^{(k)}$ atoms
at level $\ell$. Since the total number of atoms is conserved, $\sum_{\ell=0,1}N_{\ell}^{(k)}=N$. The one-level eigenstates can be expressed as $|\phi^{(K_{\ell}^{(k)})}\rangle=\sum
c_{i\ell}^{(k)} |n_{L\ell}^{(k)},n_{R\ell}^{(k)}\rangle $ with
\begin{equation}\label{eq:cilmlcoeff}
  c_{i\ell}^{(k)} \!\!= \! a\!_{K_{\ell}^{(k)}}\!\!\left(\!N_{\ell}^{(k)}\!\right)\!
  h\!_{K_{\ell}^{(k)}}\!\!\left(\!n_{L\ell}^{(k)}\!\left|N_{\ell}^{(k)}\right.\!\right)\!
  p\!\left(\!n_{L\ell}^{(k)}\!\left|N_{\ell}^{(k)}\right.\!\right)\!,
\end{equation}
where $n_{L\ell}^{(k)}$ is the occupation of the left well for the corresponding Fock vector,  $ h_{K_{\ell}^{(k)}}(n_{L\ell}^{(k)}|N_{\ell}^{(k)})$ is a $K_{\ell}^{(k)}$th order discrete Hermite polynomial,  $p(n_{L\ell}^{(k)}|N_{\ell}^{(k)})$ is the square
root of the binomial distribution,
\begin{equation}
  p\left(\!n_{L\ell}^{(k)}\!\left|N_{\ell}^{(k)}\right.\!\right)=
  \frac{1}{2^{N_{\ell}^{(k)}/2}}\sqrt{\frac{N_{\ell}^{(k)}!}{n_{L\ell}!(N_{\ell}^{(k)}-n_{L\ell}^{(k)})!}}
\end{equation}
and $a_{K_{\ell}^{(k)}}(N_{\ell}^{(k)})$ is the normalization factor
\begin{equation}
  a_{K_{\ell}^{(k)}}(N_{\ell}^{(k)}) = \sqrt{\frac{(N_{\ell}^{(k)}-K_{\ell}^{(k)})!}{N_{\ell}^{(k)}!\,K_{\ell}^{(k)}!}}\,.
\end{equation}
The eigenenergies
$\varepsilon^{(k)}$ can be expressed in terms of the one-level eigenenergies as \begin{equation}\label{eq:eigenval}
  \varepsilon^{(k)} = \sum_{\ell}\varepsilon_{(k) \ell}\,,
\end{equation}
where
 \begin{equation}
\label{eq:eigenv_OSC}
  \varepsilon_{(k), \ell } = -J_{\ell }\left(N_{\ell}^{(k)}-2K_{\ell}^{(k)}\right) + E_{\ell} N_{\ell}^{(k)}\,.
\end{equation}
The one-level Hamiltonians, expressed in the Fock basis is, in the non-interacting limit,  a harmonic oscillator potential
truncated at hard walls. The analytical expressions of 
the
Fock-space amplitudes $ c_{i\ell}^{(k)} $ resemble the probability amplitudes of observing the particle in position $x$ for the 1D harmonic oscillator problem, obtained from the expression of its eigenfunctions in the position representation. Moreover, in both cases the eigenvalues $\varepsilon_{k\ell }$ are linear in $K_{\ell}^{(k)}$.  The analogy can be established between occupation of the wells in the double well and the positions in the Harmonic oscillator problem. Since  one variable is discrete while the other is continuous the analogy is valid for the limit of infinite particles. Let us show this for the one level approach, which we obtain considering  $\hat{H}=\hat{H}_0$, where $\hat{H}_0$  is given by Eq.~(\ref{eq:HOHam}). In the Heisenberg picture the evolution of the operators $\hat{b}_{j 0}$ for $j=L,R$ is given by  \begin{equation}
 i\frac{d \hat{b}_{j 0}}{d t}=[\hat{b}_{j 0},\hat{H}]=-J \hat{b}_{j' 0}\,,
 \label{eq:HeisPict}
\end{equation}
 with  $j'\neq j$, and where we considered $E_0=0$ for simplicity. In the following we omit the level index $\ell$ because we are considering only the one level approximation. For $N\rightarrow \infty$, we can approach the operators by $c$-numbers $\alpha_j=\sqrt{N_j}e^{i\theta_j}$. Substituting we obtain  the two corresponding equations of motion which in turn can be obtained from the pendulum Hamiltonian
\begin{equation}
 \mathcal{H}=-2J\sqrt{N_L\,N_{R}}\cos(\theta_L-\theta_{R})=-J\,N\sqrt{1-z^2}\cos(\theta)\,,
\end{equation}
where $N=N_L+N_R$, $ z=(N_L-N_R)/N$, $\theta=\theta_L-\theta_{R}$.  The previous approach is widely known to be extended to the small interacting regime, where it gives a Hamiltonian analogous to that of a non-rigid pendulum.
The study of the dynamics of such a Hamiltonian permits one to predict macroscopic quantum tunneling and self-trapping within the semiclassical approach~\cite{1986JavanainenPRL,1997MilburnPRA,1997SmerziPRL}. In our case, this can be extended to a harmonic oscillator Hamiltonian, where the angular variables can take any value. The analogy here is with the harmonic oscillator truncated at hard walls, because the number of particles is finite. Due to  this analogy, the eigenstates (\ref{eq:HO_states}) are said to be harmonic-oscillator-like. In Sec.~\ref{sec:bounds} we  obtain numerically the eigenvectors and eigenvalues for small interactions, showing that for non-zero interactions the eigenvectors and eigenvalues closely resemble the ones characterized for the non-interacting regime.

\subsection{Fock Regime}
\label{sec:MS}

In the extreme Fock regime, or the infinite-barrier limit, $J_0=J_1=0$. In this regime, the coefficient $ U_{01 } $ cannot be
neglected. Then, the eigenvectors of Hamiltonian~(\ref{eq:two-level}) are  Fock vectors, if we further assume that $ 2\triangle E\gg N^2 U_0$. The
latter assumption is needed because the term  $\hat{b}_{jl}^{\dagger}\hat{b}_{jl}^{\dagger}\hat{b}_{jl'}\hat{b}_{jl'}$ couples Fock vectors with
atoms in different levels. As demonstrated in Sec.~\ref{sec:bounds}, this coupling is small under that assumption. In this limit, the eigenvalues are
\begin{align}
\label{eq:eigenHB}
\varepsilon^{(k)}&= \sum_{\ell}\Bigg\{E_{\ell} N_{\ell }^{(k)}+U_{\ell }\left[2\left(n_{L\ell }^{(k)}-\frac{N_{\ell
}^{(k)}}{2}\right)^{2}\right.\nonumber\\ &+\left.N_{\ell }^{(k)}\left(\frac{N_{\ell }^{(k)}}{2}-1\right)\right]\Bigg\}+U_{01}\sum_{j\ell}2\,n_{j\ell
}^{(k)}n_{j\ell' }^{(k)}\,. \end{align}

Here, the spectra reflects the main symmetry in the problem: that of discrete rotations of an angle $\pi$ around an axis perpendicular to the $x-y$ plane that intersects this plane in the origin. This in turn means that the eigenstates, which are the Fock vectors $ |n_{L0},n_{R0}\rangle|n_{L1},n_{R1}\rangle$, are degenerate in pairs with those obtained for $L \to R$.
Small but non-negligible hopping terms,   $J_0 \ll U_0$ and $J_1 \ll U_0$, break this degeneracy, as can be shown by non-degenerate high-order perturbation theory. Then, in the small tunneling limit, that is, in the Fock regime, the eigenvectors are quasi-degenerate symmetric and antisymmetric superpositions of those pairs of Fock vectors that are degenerate in the infinite barrier case.  For the particular cases in which all atoms occupy the same level the eigenstates are  direct products of one-level vectors
$|\phi_{\ell}^{\pm}\rangle|0,0\rangle_{\ell'} $ with \begin{align}
  & |\phi_{\ell}^{\pm}\rangle
  \equiv\frac{e^{i\theta_0}}{\sqrt{2}}\left[\frac{1}{\sqrt{\nu!(N-\nu)!}}
    \left(\bhat{L}{\ell}{\dagger}\right)^{\nu}
    \left(\bhat{R}{\ell}{\dagger}\right)^{N-\nu}\right.\nonumber\\
& \pm
   \left.\frac{1}{\sqrt{\nu!(N-\nu)!}}
   \left(\bhat{L}{\ell}{\dagger}\right)^{N-\nu}
   \left(\bhat{R}{\ell}{\dagger}\right)^{\nu}
  \right]|0\rangle\,,
\end{align} for $0\leq \nu < N/2$. These are  MS states in which $\nu$ and $N-\nu$ atoms simultaneously occupy the $\ell$th energy level of both
wells;  although the expression might appear complicated, in fact it is just a two-state approximation.  We have neglected terms on the order of $(J_{\ell}/U_{\ell})^{N-2\nu}$ and smaller.    Here $ \theta_0$ is the usual arbitrary phase
associated with vectors in a Hilbert space. We will set $ \theta_0=0$ for the rest of this Article.

The special case $\nu=0$ represents an
 extreme MS state, or \textit{NOON state}, in which all $N$ atoms simultaneously occupy the left and right wells, which can be written also as   $ (1/\sqrt{2})\left(|N,0\rangle|0,0\rangle\pm |0,N\rangle|0,0\rangle\right)$. These symmetric $(+)$ or
antisymmetric $(-)$ MS states are nearly degenerate, with  a splitting $\Delta\varepsilon_{\ell}(\nu)$ between the states $|\phi_{\ell}^{+}\rangle$ and $|\phi_{\ell}^{-}\rangle$ given by

\begin{equation}
  \Delta\varepsilon_{\ell}(\nu) =
  \frac{4U_{\ell }[J_{\ell }/(2U_{\ell})]^{N-2\nu}(N-\nu)!}
  {\nu![(N-2\nu-1)!]^2}\,,
\end{equation}
 up to $(N-2\nu)$th order in $J_{\ell}/U_{\ell}$. General MS states are symmetric/antisymmetric superpositions of Fock vectors with
atoms in both levels, that is, $|\nu,N_{0}-\nu\rangle|\nu',N_{1}-\nu'\rangle $ and  $|N_{0}-\nu,\nu\rangle|N_{1}-\nu',\nu'\rangle $, with splittings
that are proportional to $(J_0/U_0)^{N_0-2\nu} $ $(J_1/U_1)^{N_1-2\nu'}$.

Since  $J_1>J_0$, it is also possible that $J_0 \ll U_0$ but $J_1 > U_0$;  this is the mixed regime. The atoms in the lower level behave as in the Josephson regime while the
ones in the excited level behave as in the Fock regime. We will show numerical examples of  the mixed regime in the following section.

\section{Characterization of the Bounds of the Model, Regimes, and Crossings}
\label{sec:bounds}

\subsection{Limits of the Two-level Model}
\label{sec:limits}

Let us find criteria in terms of the characteristic parameters of the problem to describe  the limits of all regimes, the occurrence of shadows of the
 MS states, crossings,  and the limits of applicability of the Hamiltonian, Eq.~(\ref{eq:two-level}).  We begin by seeking criteria of validity for the model. On the one hand,  the consideration of  quantum tunneling associated with a second level of energies is meaningful if $V_0>E_1$, as  evident from Fig.~\ref{fig0}(a).  If the eigenfunctions are approximated by the eigenfunctions of the harmonic oscillator, then this criterion can be approximated by  $V_0> (3/2)\hbar\omega$.    Secondly, the diluteness condition given by
Eq.~(\ref{eq:diluteness}) can be written in three dimensions as  $N^{1/3}U_0^{3D}/\Delta E\ll 1$~\cite{2011GarciaMarchPRA}. In 1D,
this becomes
\begin{equation}
 \frac{\omega N^{1/3}U_0^{3D}}{\omega_{\perp}\Delta E} \ll 1\,,
\end{equation}
which, contrary to the 3D case, accounts also for the geometrical compression in one of the dimensions through the term $\omega/\omega_{\perp}$. Here, it is also important to consider the restrictions on $\omega_{\perp}$  discussed at the beginning of   Sec.~\ref{sec:1DLMG}. Finally,  a last limit of the model is obtained when the coupling between levels is big enough   to require more levels to characterize the eigenvectors.  We obtain below the criterion delimiting when this coupling is too big.  If this criterion is  not fulfilled, more levels are required in the approximation or one should use the MCTDH method~\cite{2005MasielloPRA,2006StreltsovPRA,2008AlonPRA},  depending on numerical efficiency requirements and spatial dimensionality. Other methods such as matrix product state algorithms, for example, time-evolving block decimation, can also of utility in this regime~\cite{2011schollwoeckAP}.

\subsection{Discussion of the Different Regimes}
\label{sec:shadows}

\begin{figure}
\begin{tabular}{c}
 \vspace{-0.cm}(a)\\
 \includegraphics[trim =43mm 2mm 29mm 10mm, clip, width=8.5cm]{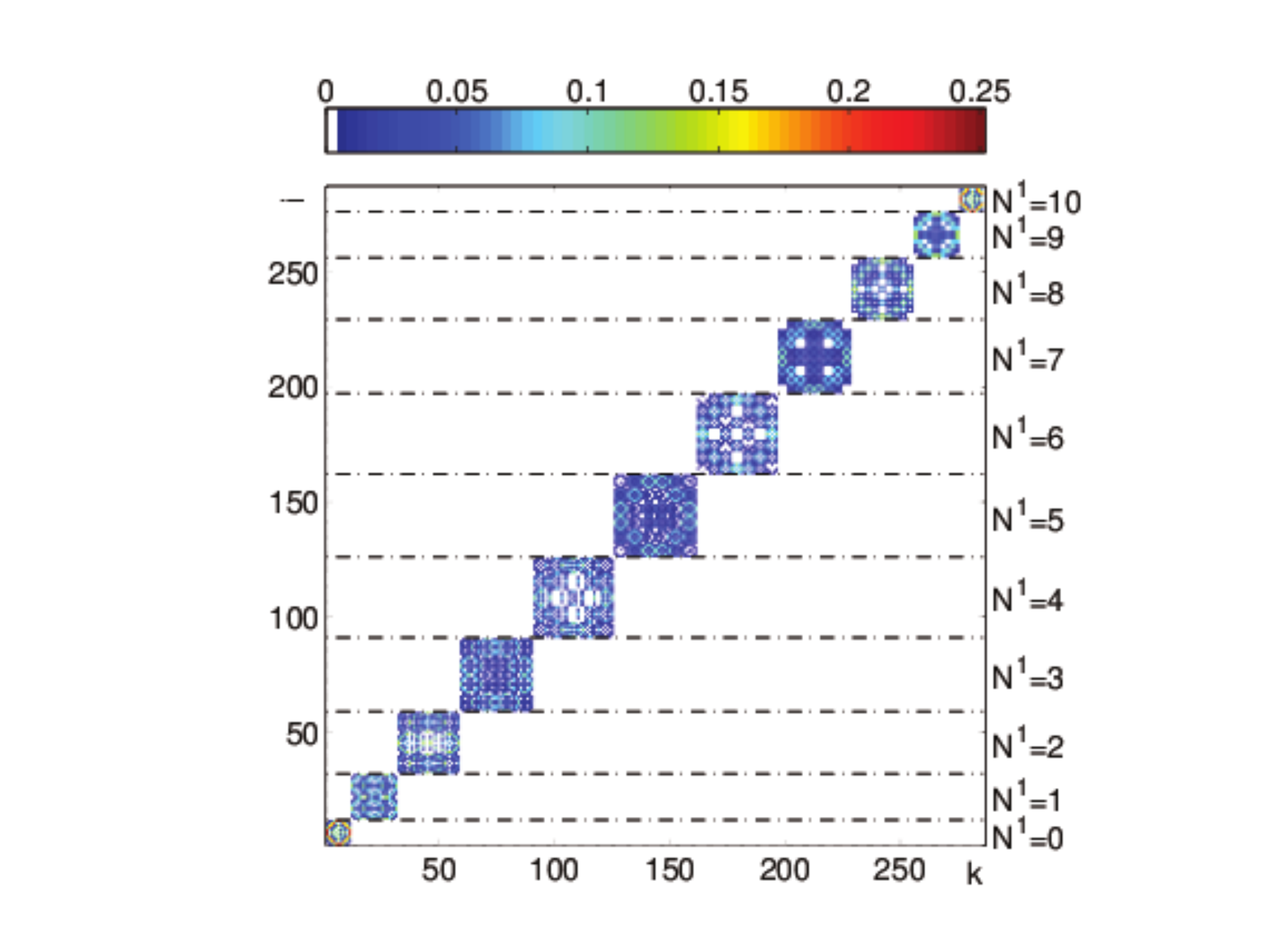}\vspace{-0.6cm} \\
\begin{tabular}{cc}
(b) & (c) \\
\includegraphics[trim =32mm 2mm 30mm 10mm, clip, width=3.6cm]{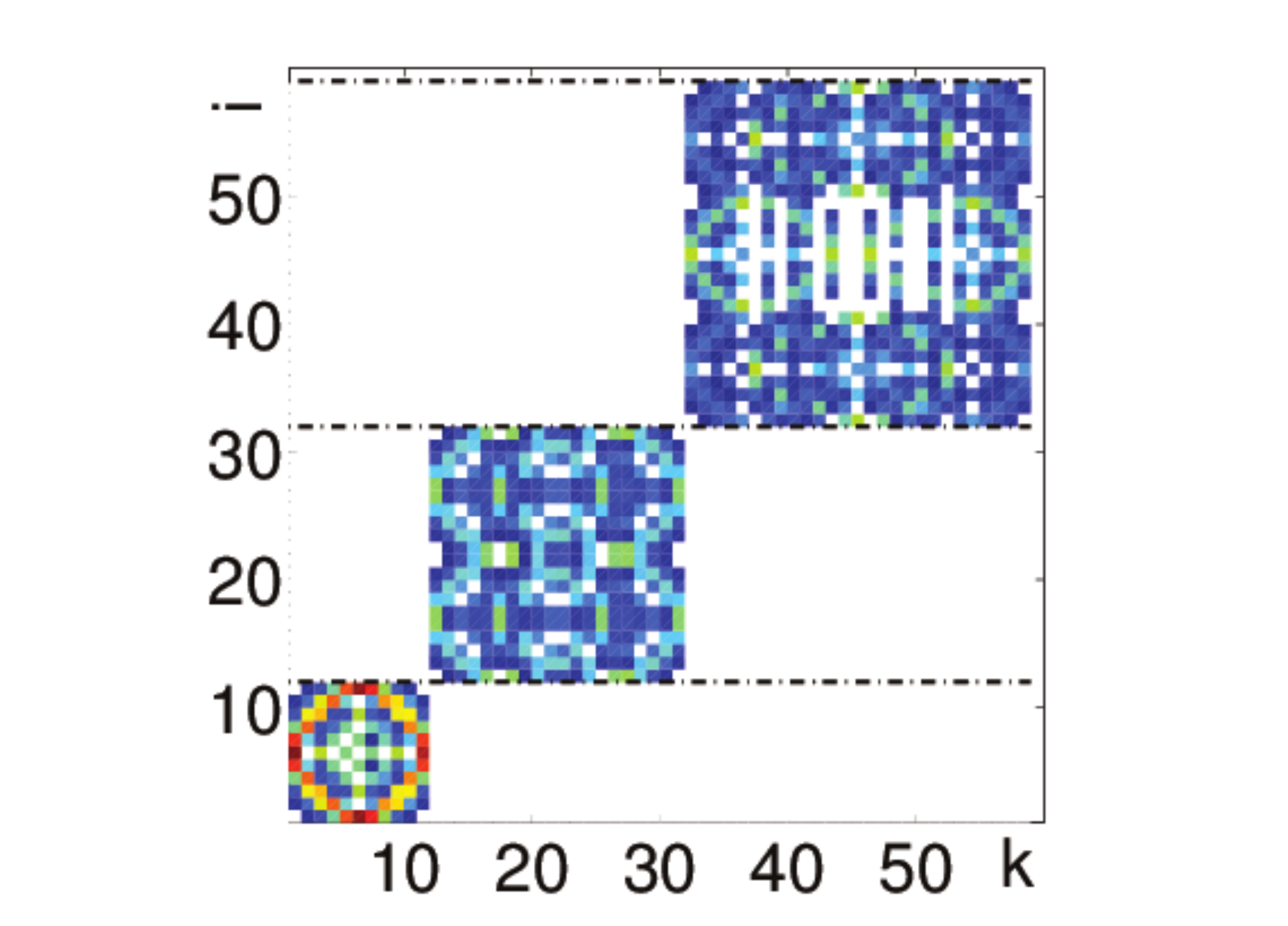} &  \includegraphics[trim =25mm 2mm 30mm 10mm, clip,width=3.8cm]{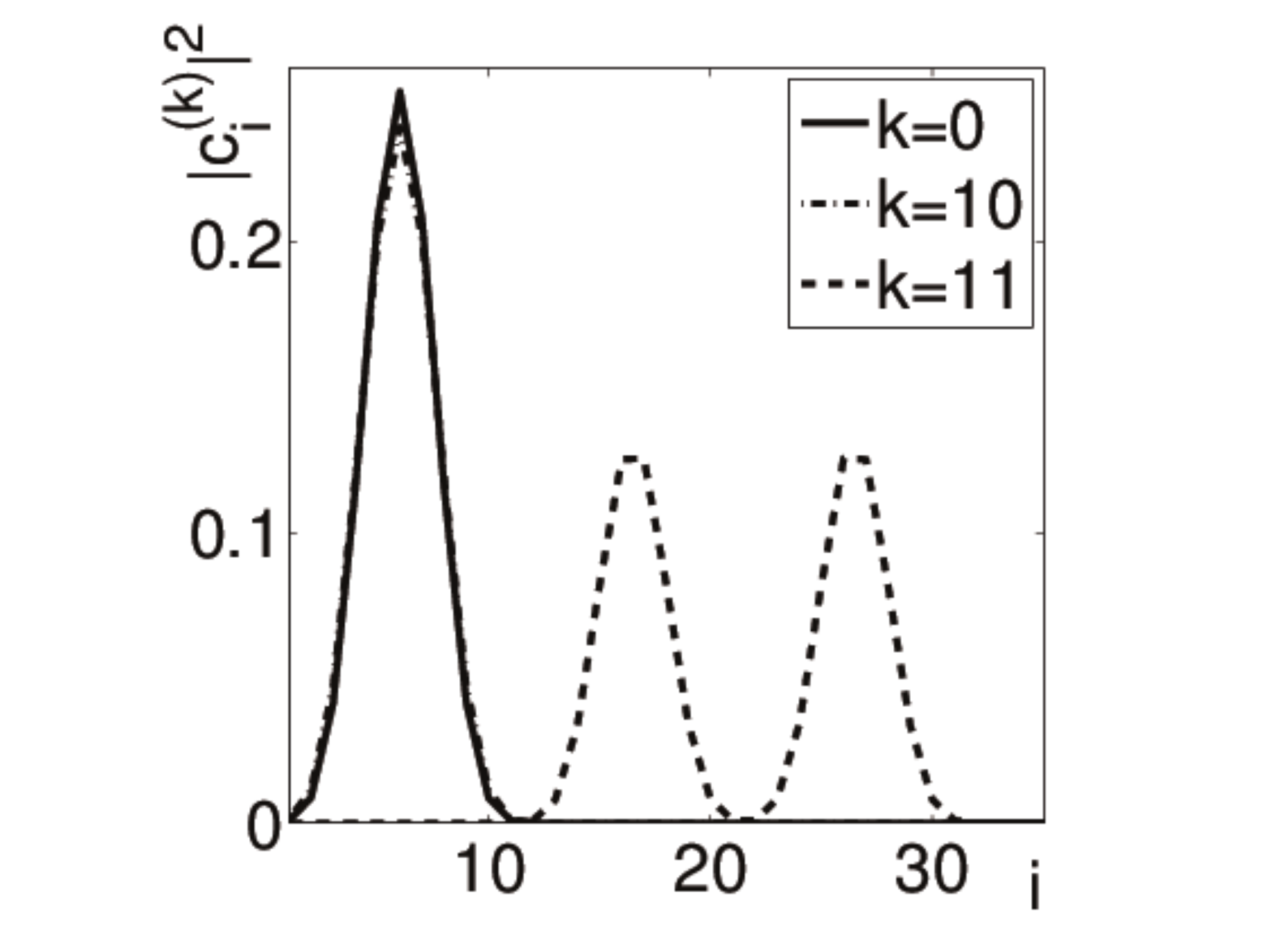}\vspace{-0.2cm} \\
\end{tabular}\\
(d)\\
\includegraphics[trim =35mm 2mm 30mm 10mm, clip,width=6cm]{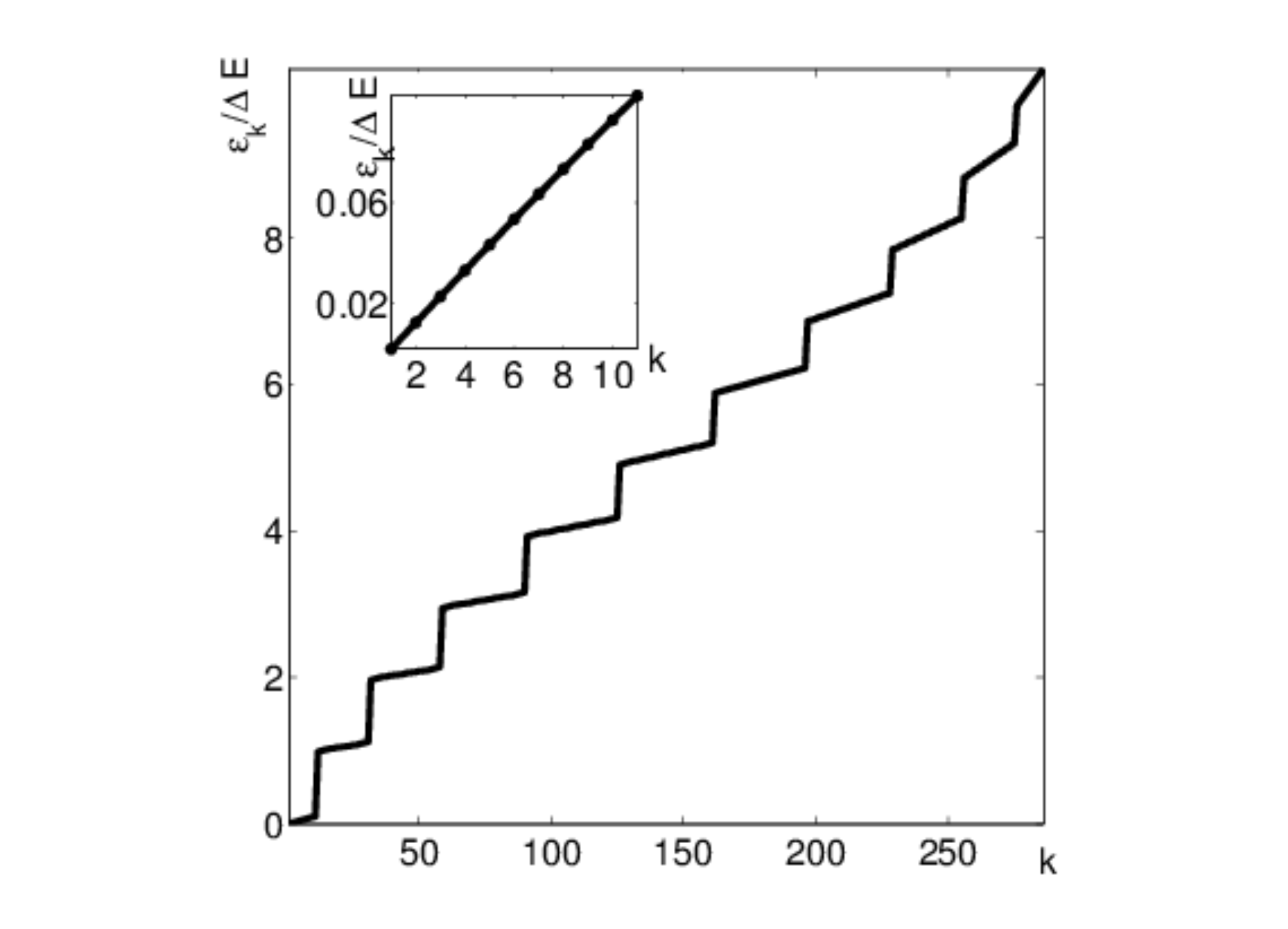}\vspace{-0.3cm}
\end{tabular}
 \vspace{-0.4cm}
 \caption{(Color online)  {\it  Josephson regime example.}
(a)  Fock-space amplitudes for all the eigenvectors, which show the harmonic oscillator behavior. (b) Zoom of panel (a). (c) Fock-space amplitudes for the ground state, $(N+1)$th, and $(N+2)$th excited eigenstates.  ((d) Eigenvalues as a function of Fock state index.  Inset: lowest level approximation, first N+1 states only.  The energy increases linearly, with characteristic jumps occurring each time a particle enters the excited level.    \label{fig1}} \end{figure}

\begin{figure}

\begin{tabular}{c}
 \vspace{-0.cm}(a)\\
 \includegraphics[trim =43mm 2mm 29mm 10mm, clip,width=\columnwidth]{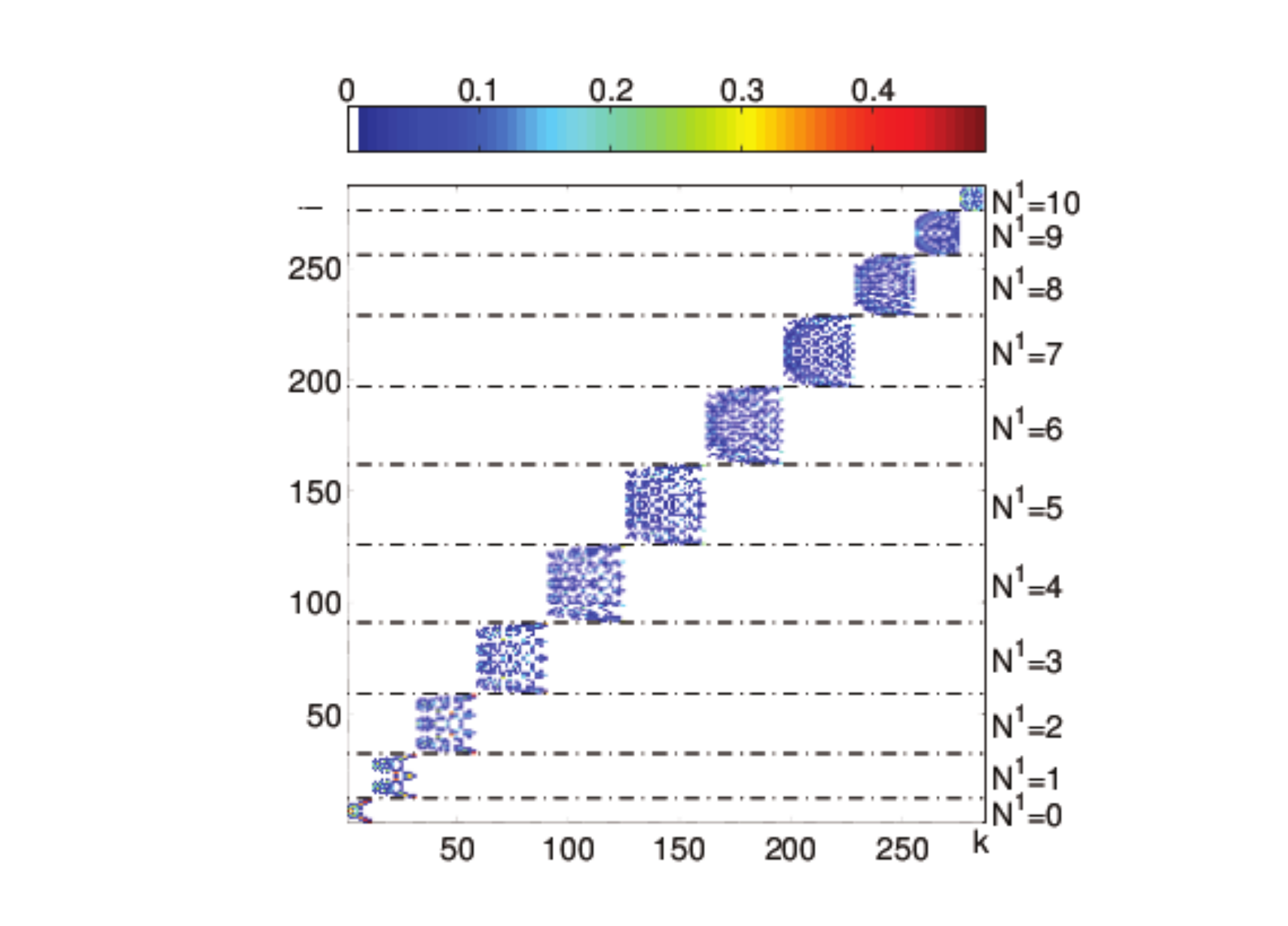}\vspace{-0.6cm} \\
\begin{tabular}{cc}
(b) & (c) \\
\includegraphics[trim =32mm 2mm 30mm 10mm, clip, width=3.6cm]{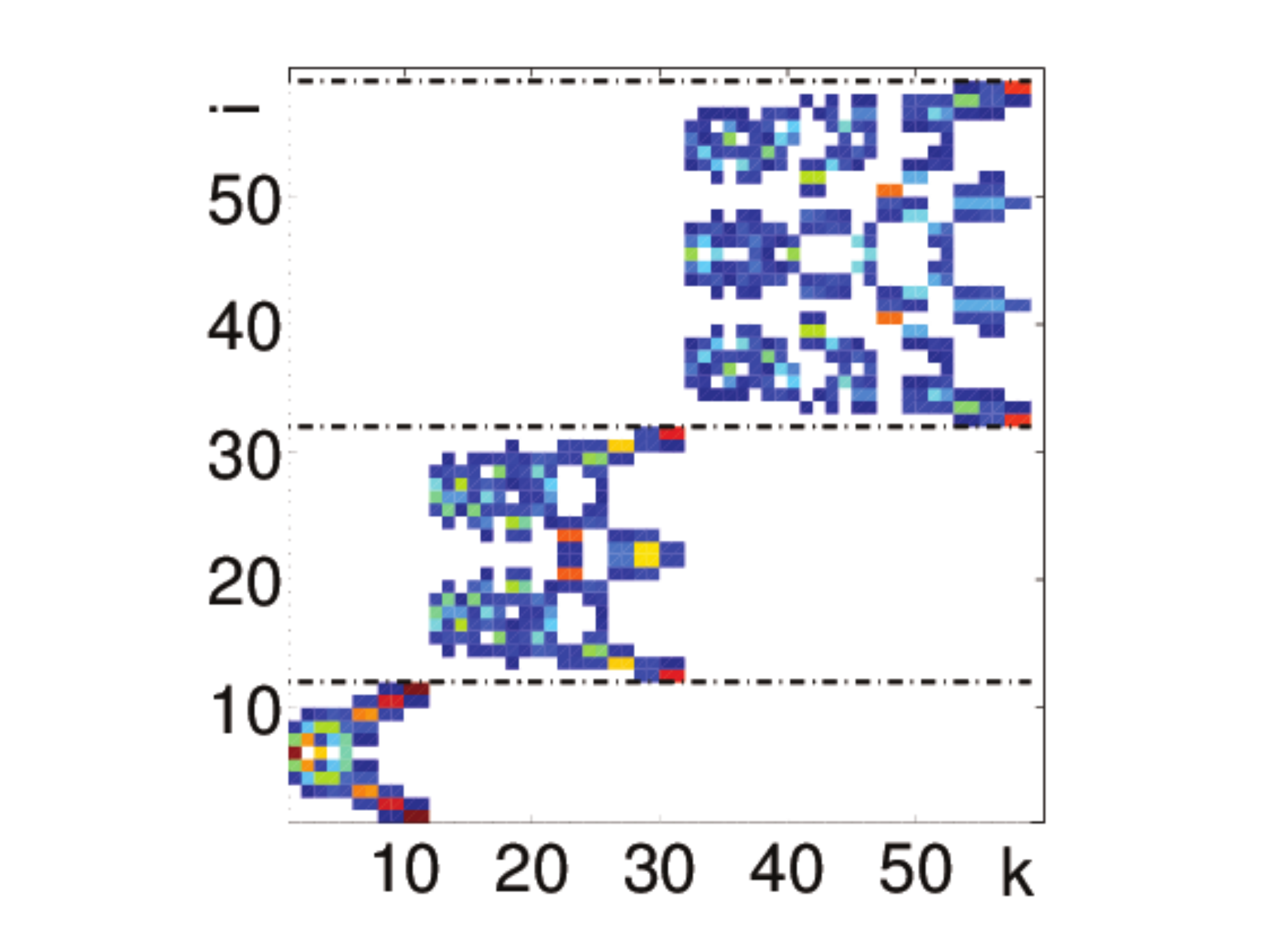} &  \includegraphics[trim =29mm 2mm 30mm 10mm, clip, width=3.7cm]{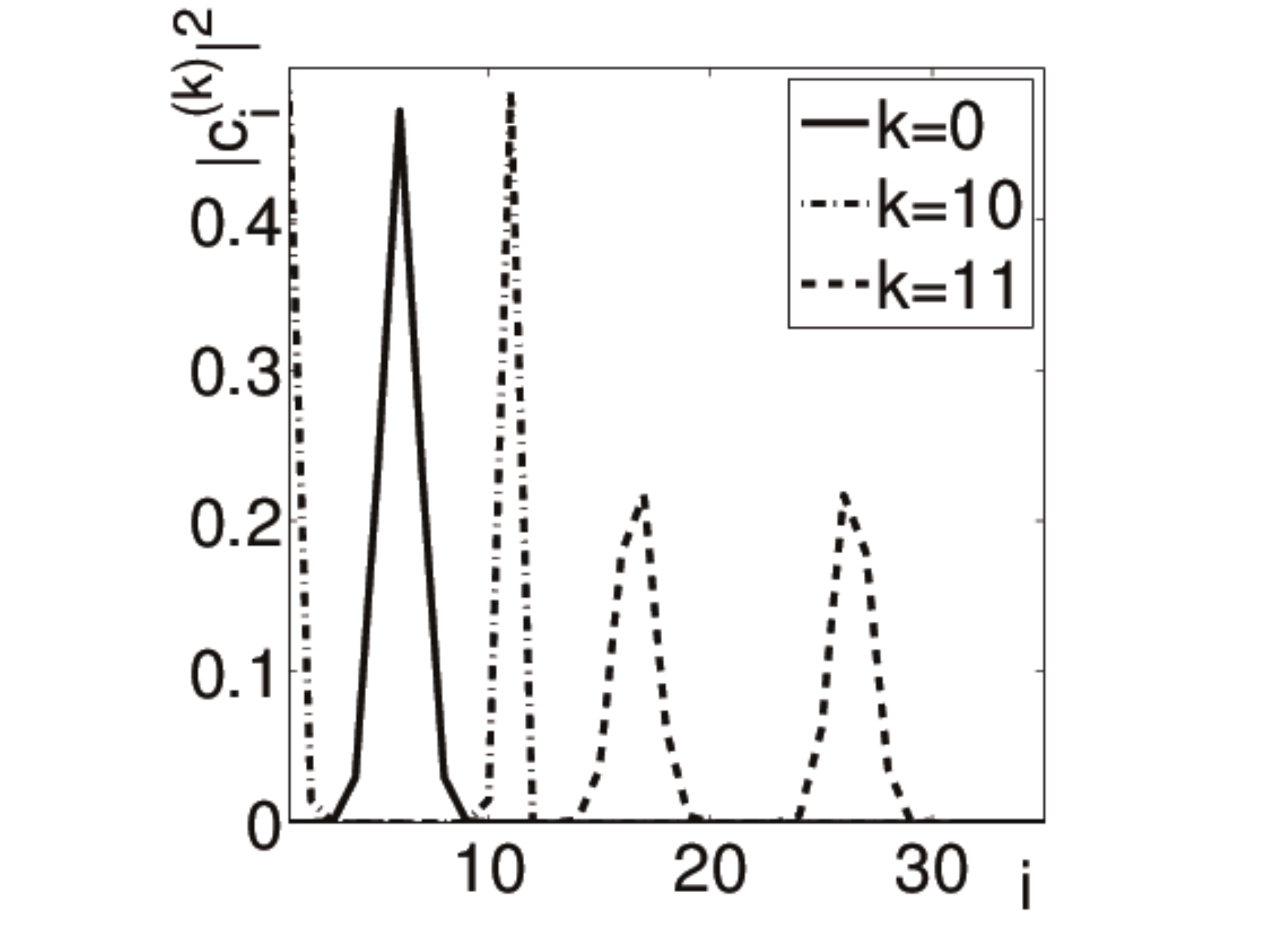}\vspace{-0.2cm} \\
\end{tabular}\\
(d)\\
\includegraphics[trim =35mm 2mm 30mm 10mm, clip,width=6cm]{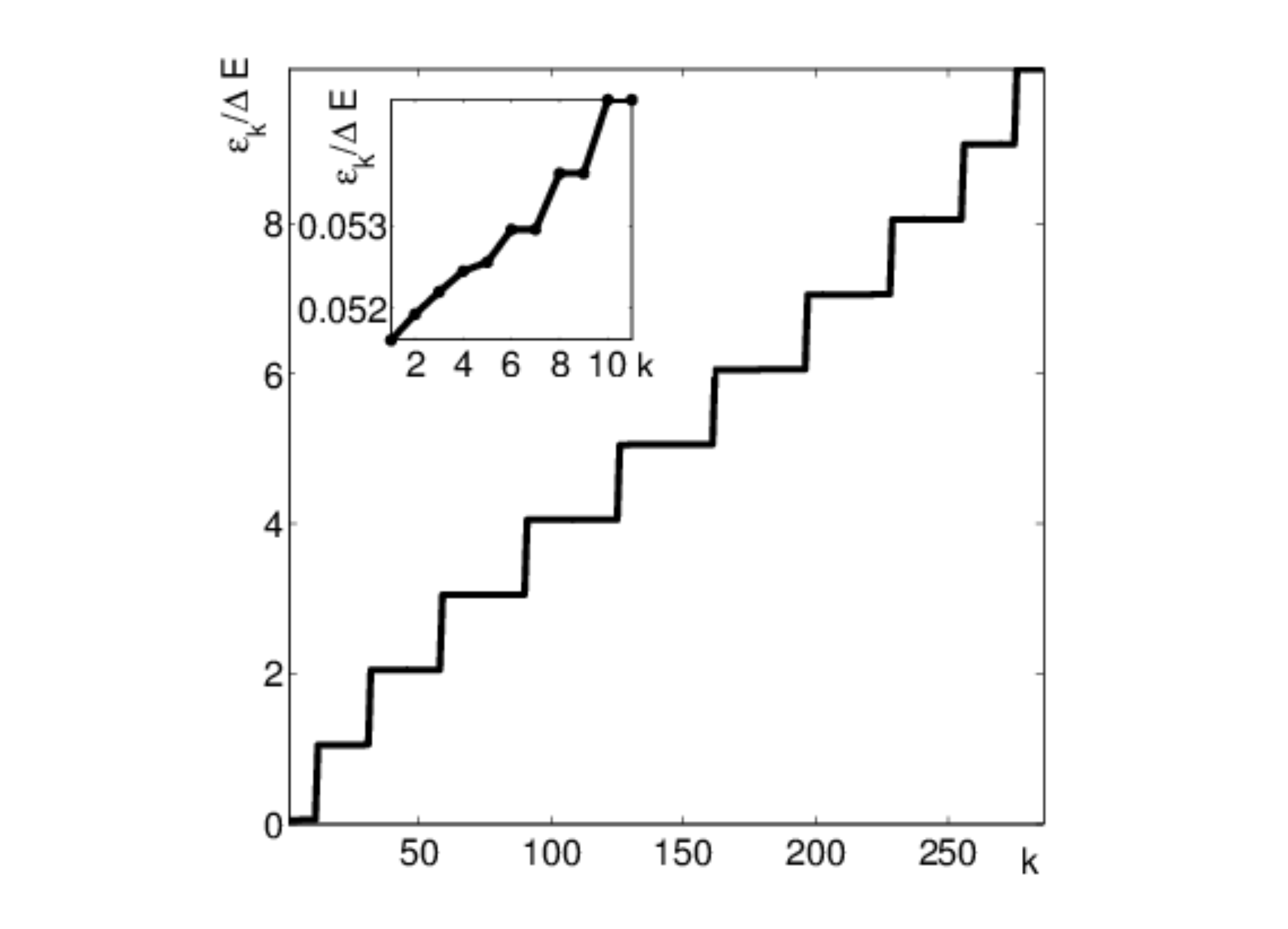}\vspace{-0.3cm}
\end{tabular}
 \vspace{-0.4cm}
   \caption{(Color online)  {\it Intermediate regime example.}
 Panel layout as in Fig.~\ref{fig1}. The lowest excited $N+1$ eigenvectors  show no occupation of the lower level. Among these, the lowest excited 5 show harmonic oscillator behavior, while the next 6  eigenvectors behave as in the Fock regime.   The eigenvalues show also this mixed HO-MS behavior.  Notice that the $N+1$ more excited eigenstates of the spectra are harmonic oscillator like.      \label{fig2}} \end{figure}

\begin{figure}

\begin{tabular}{c}
 \vspace{-0.cm}(a)\\
 \includegraphics[trim =43mm 2mm 30mm 10mm, clip, width=\columnwidth]{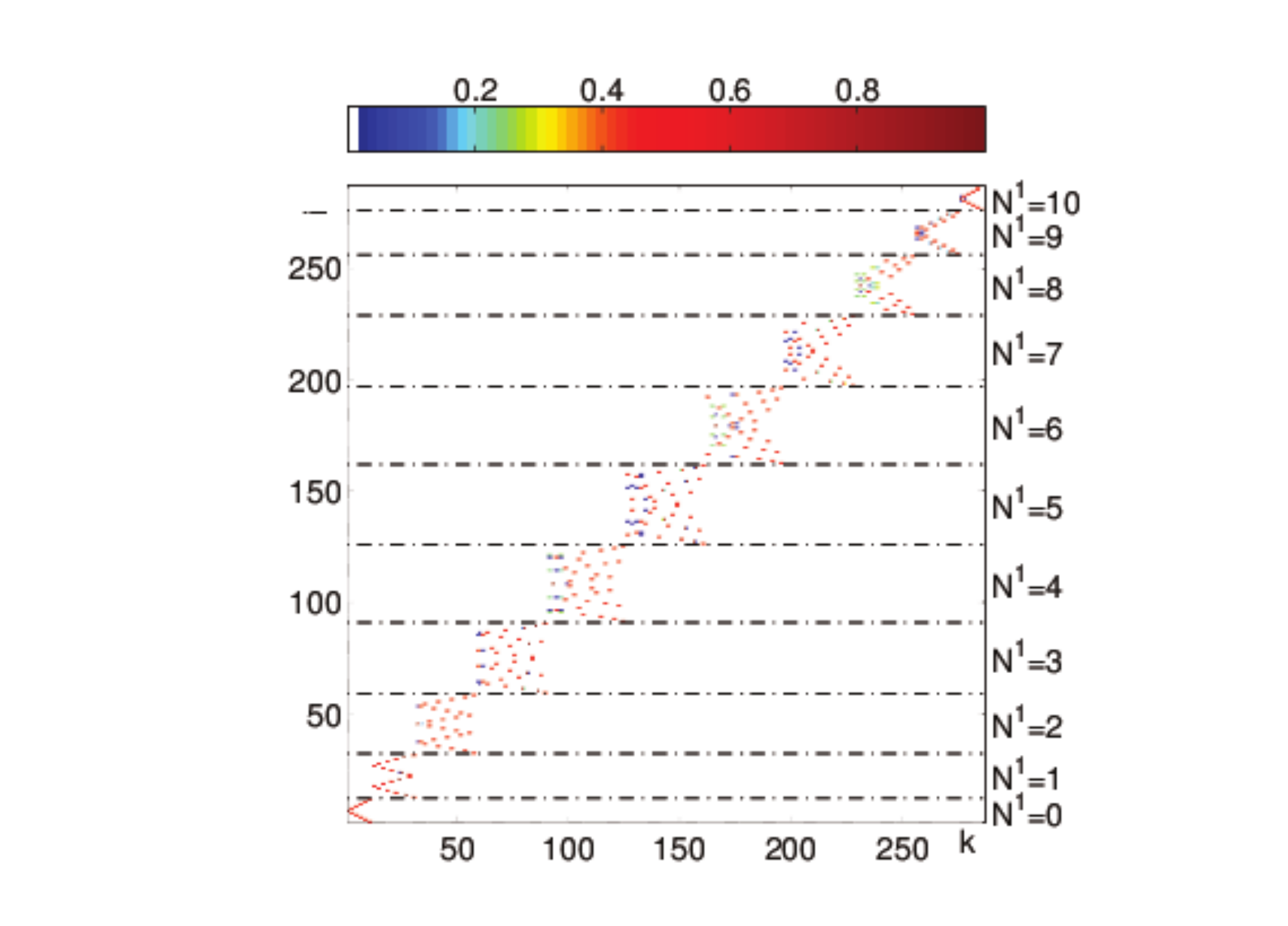}\vspace{-0.6cm} \\
\begin{tabular}{cc}
(b) & (c) \\
\includegraphics[trim =32mm 2mm 30mm 10mm, clip, width=3.6cm]{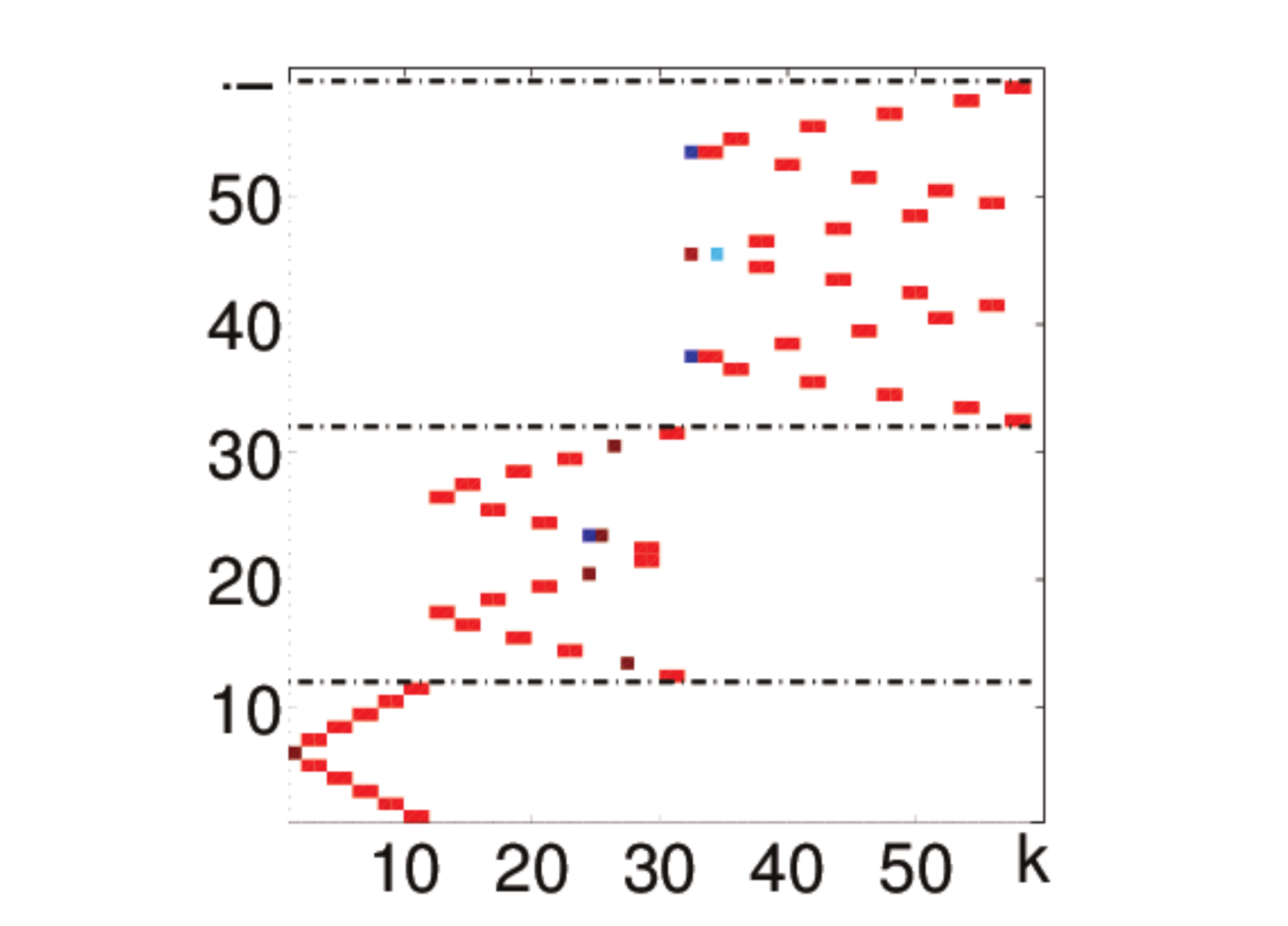} &  \includegraphics[trim =24mm 2mm 30mm 10mm, clip, width=3.8cm]{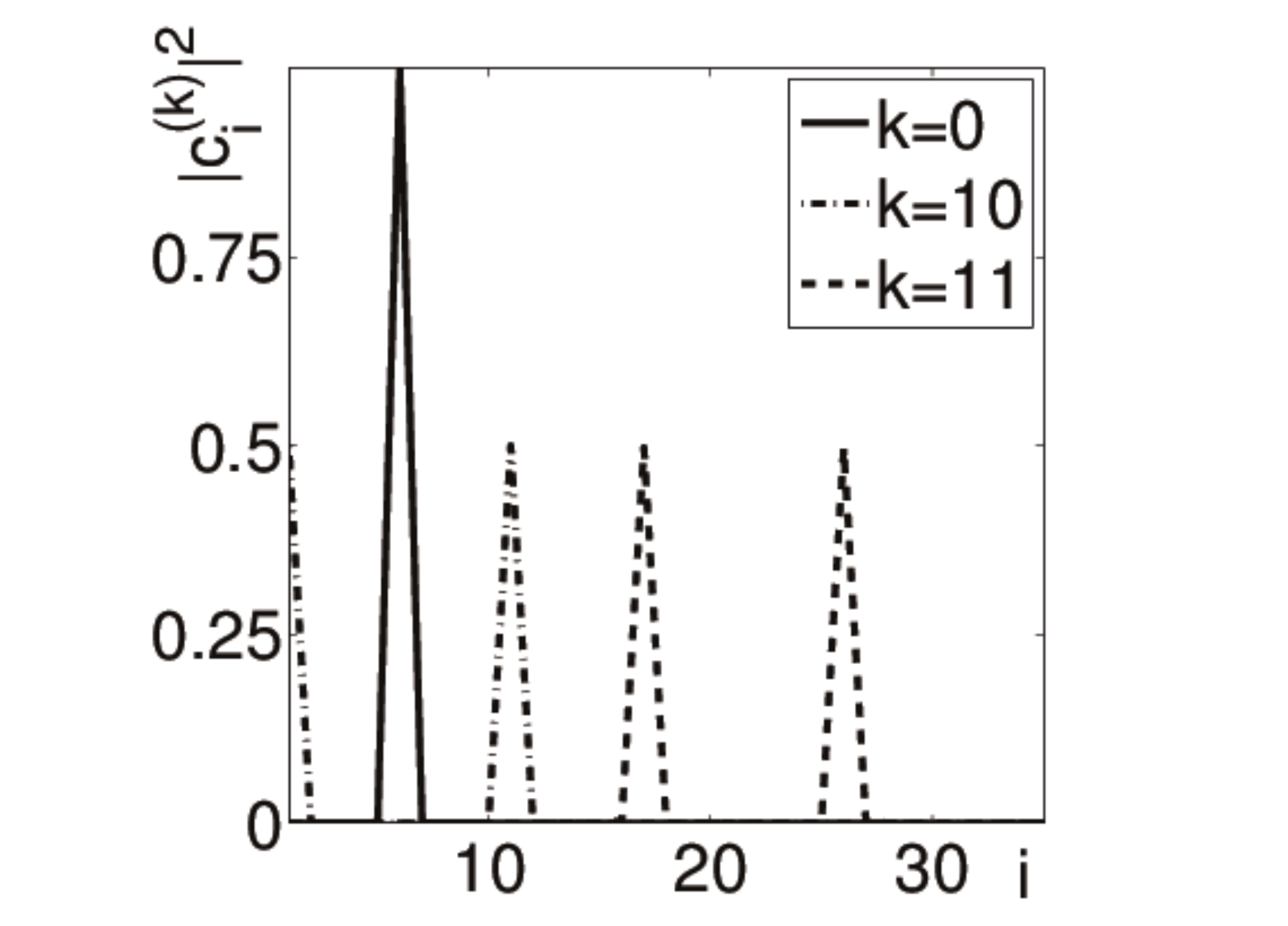}\vspace{-0.2cm} \\
\end{tabular}\\
(d)\\
\includegraphics[trim =35mm 2mm 30mm 10mm, clip,width=6cm]{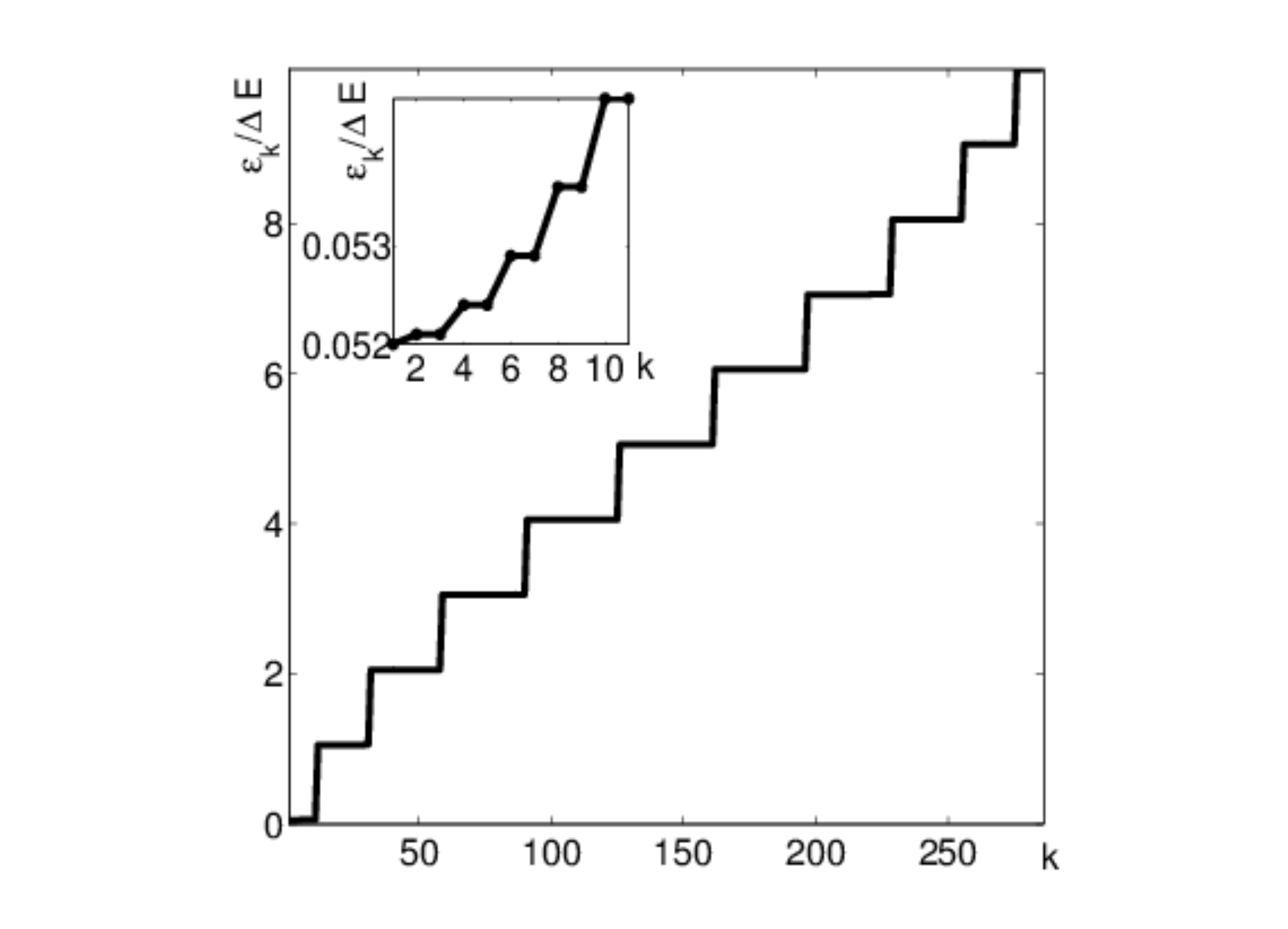}\vspace{-0.3cm}
\end{tabular}
 \vspace{-0.4cm}
\caption{(Color online)  {\it A typical example in the Fock regime.} Panel layout as in Fig.~\ref{fig1}. Here, the  eigenstates appear in quasidegenerate pairs of MS states.   \label{fig3}} \end{figure}

\begin{table}[htp]
 \centering
\begin{tabular}{c|c|c}
Fock index $i$ & $N_0$ & $N_1$\\
\hline
1 to $N+1$ & $N$ & 0\\
\hline
$N+2$ to $3 N+2$ & $N-1$ & 1\\
\hline
$3 N+3$ to & & \\
$3 N+3+3(N-1)$ & $N-2$ & 2
\end{tabular}
 \caption{Fock indices for the the Fock vectors up to the $(3 N+3+3(N-1))$-th resulting from the ordering given in Eq.~(\ref{Eq:Fockindexordering}). These include all  combinations with up to two particles excited.}
 \label{tab}
\end{table}

The Josephson and Fock regimes are characterized by the criteria discussed above, Eqs.~(\ref{eq:CA}) and (\ref{eq:CB}). Next we will show some numerical examples of the direct diagonalization of the Hamiltonian~(\ref{eq:two-level}) for different sets of parameters, corresponding to different regimes.  In Fig.~\ref{fig1}   we show an example of the Josephson regime, $U_1=U_0/2$, $U_{01}=(3/4) U_0$, $\zeta_0=10^2$, $J_1=5J_0$, $\Delta E =
\sigma N U_0$, with $N=10$ atoms and $\sigma=2\cdot10^3$.   In Fig.~\ref{fig1}(a) we show all the Fock-space amplitudes $|c_i|^2$ for all the eigenvectors, while Fig.~\ref{fig1}(b) shows a zoom of these for the lowest excited $3 N+3(N-1)+1$ states. The first $N+1$ are eigenstates with no coupling to Fock vectors with atoms in the excited level. Then, the following $2N$ are eigenvectors where the Fock vectors show $N-1$ atoms in the lower level and one atom in the excited level. Finally, the following   $3(N-1)$ show two atoms excited and $N-2$ in the lower level of eigenenergies (see table~\ref{tab}).   In Fig.~\ref{fig1}(c) we show the Fock-space amplitudes $|c_i|^2$ for the ground state, $(N+1)$th, and $(N+2)$th excited eigenstates. These two figures show the HO behavior discussed in Sec.~\ref{subsec:HO}.    In Fig.~\ref{fig1}(d)   we
represent all the eigenvalues while in the inset in this figure we represent a zoom for the lowest excited $N+1$ eigenvectors, where the linear behavior described
by Eq.~(\ref{eq:eigenv_OSC}) is shown. Every step in the values of the eigenvalues in  Fig.~\ref{fig1}(d) occurs when the corresponding eigenstate is the superposition of Fock vectors with one more atom occupying  the second level of energies. The increase in the slope in
every step is due to the fact that $J_1>J_0$. Indeed, for the last $N+1$ eigenvectors, for which all atoms are excited to the higher level, the slope
is $J_1$.

If we reduce $\zeta_0$ to one, we cannot assume that the parameters correspond clearly to the Fock or to the Josephson regime, as
represented in  Fig.~\ref{fig2}(a). The Fock-space amplitudes  $|c_i|^2$ of the $N+1$  lowest excited eigenvectors show different behavior for the ground state
and for the $(N+1)$th excited state, as represented in Fig.~\ref{fig2}(b) and (c). The ground state is still a HO eigenstate, while the  $(N+1)$th excitation
is a NOON-like state, as corresponding  to the Fock regime.  Fig.~\ref{fig2}(d) shows a different behavior than in the previous case.
Particularly, as shown in  the inset in  Fig.~\ref{fig2}(d), the eigenvalues behave linearly for the five lowest excited eigenvectors. Then, the next six appear in quasidegenerate pairs with small splittings, as corresponds for the Fock regime.  Notice that, since $J_1> J_0$, the less excited $N+1$ eigenvectors behave
as HO eigenstates. Then in this mixed regime, states with atoms occupying only the lower level can behave as in the Fock regime, while states with  atoms occupying the
excited level behave as in the Josephson one.

If we decrease $\zeta$ further, to $\zeta=10^{-2}$, we enter clearly in the Fock regime, as
represented in Fig.~\ref{fig3}. Now, the eigenstates are  MS states, as shown in  Fig.~\ref{fig3}(b) and (c), while the ground state is sharply peaked in $
|N/2,N/2\rangle |0 ,0\rangle $.   Also, Fig.~\ref{fig3}(d) shows that the MS states appear in  quasidegenerate symmetric/antisymmetric pairs,
with small splittings in the eigenvalues (see also inset in this figure). Notice that, when atoms are excited, the presence of the term $2 U_{01}\sum_{j, \ell\neq\ell'} \hat{n}_{j \ell}\hat{n}_{j
\ell'}$ breaks the degeneracy in the $J_{\ell}=0$ case between the vectors  $ |N_0-\nu,\nu\rangle |N_1-\nu' ,\nu'\rangle $ and $ |N_0-\nu,\nu\rangle
|\nu',N_1-\nu' \rangle $. For small $J_{\ell}$, this in turn makes the MS states with $N_1\ne0$ appear also in pairs, and not as the superposition of
four Fock vectors. These pairs can be observed in Fig.~\ref{fig3}(b).

Finally, the last regime is characterized by the presence of shadows  of the MS states. Using perturbation theory we can show that the MS states $
|n_{L,0},n_{R,0}\rangle|0,0\rangle\pm|n_{R,0},n_{L,0}\rangle|0,0\rangle $ couple to MS states showing two excited atoms
$|n_{L,0}-2,n_{R,0}\rangle|2,0\rangle\pm|n_{R,0},n_{L,0}-2\rangle|0,2\rangle $ and
$|n_{L,0},n_{R,0}-2\rangle|0,2\rangle\pm|n_{R,0}-2,n_{L,0}\rangle|2,0\rangle $  with  coefficients
\begin{align} c_L & = U_{01}\frac{\sqrt{2 n_{L,0}(n_{L,0}-1)}}{U_0(6-4 n_{L,0})+2U_1+2\Delta E}\,,\nonumber\\ c_R & = U_{01}\frac{\sqrt{2
n_{R,0}(n_{R,0}-1)}}{U_0(6-4 n_{R,0})+2U_1+2\Delta E}\,. \end{align}
 For the most excited MS state, $n_{L,0}=N$, and
\begin{equation}
\label{eq:shad}
c_{N}  = U_{01}\frac{\sqrt{2 N(N-1)}}{U_0(6-4 N)+2U_1+2\Delta E}\,.
\end{equation}  If this coupling is not
negligible, shadows of the MS states are coupled, and the upper level cannot be neglected, as discussed below.

\subsection{Crossings of the Eigenvalues}

Let us show, for both regimes, bounds  on the one- and two-level approximations other than criterion~(\ref{eq:shad}). With our choice of indexing
states, the one-level approximation corresponds to truncating the size of the Hilbert space to $N+1$. With this definition of the one-level
approximation, the bounds we present below will describe the regime in which this truncation is valid.

\begin{figure}

\begin{tabular}{c}
 \vspace{-0.cm}(a)\\
 \includegraphics[trim =43mm 2mm 30mm 10mm, clip, width=\columnwidth]{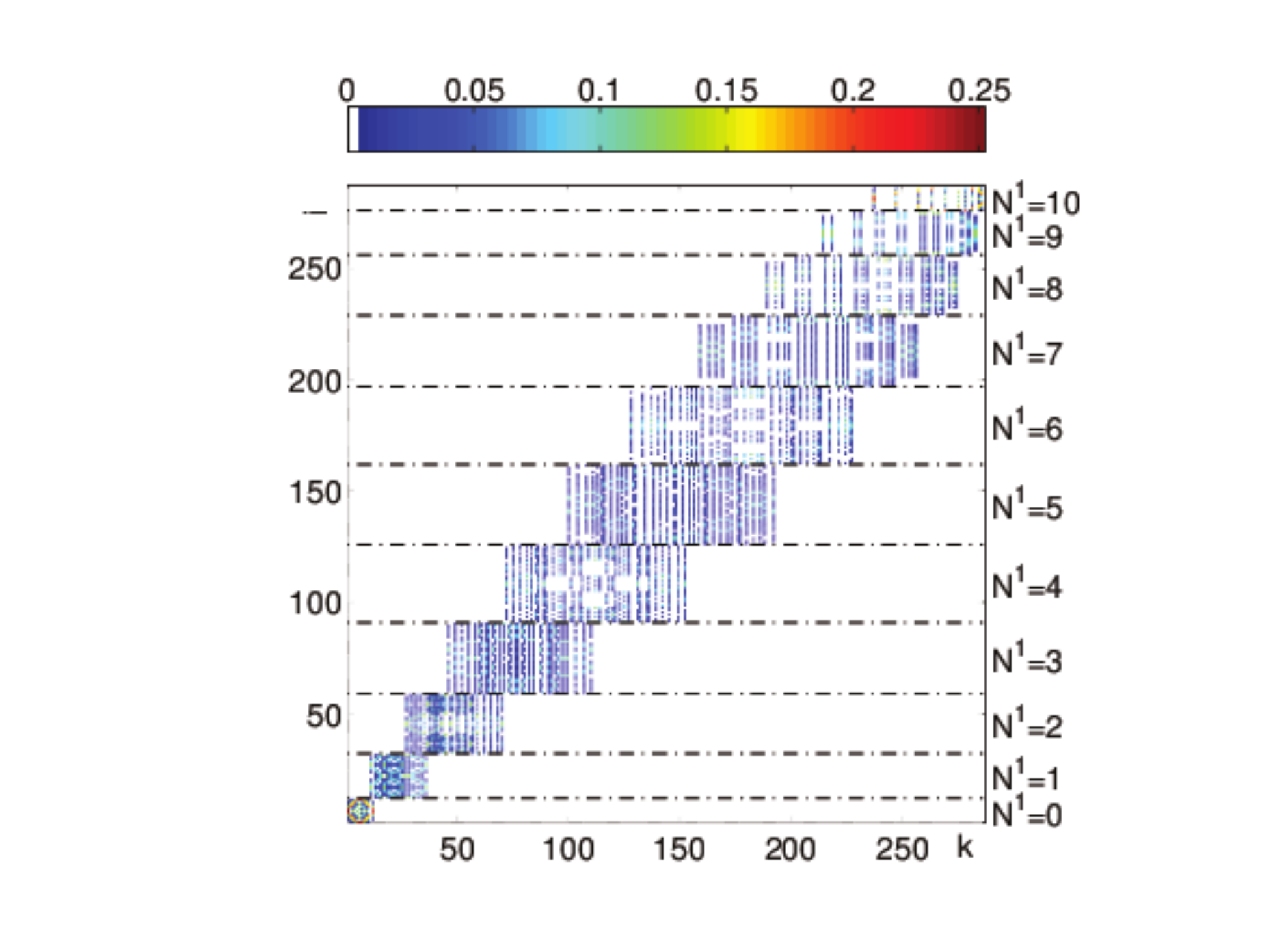}\vspace{-0.6cm}\\
\begin{tabular}{cc}
(b) & (c) \\
\includegraphics[trim =32mm 2mm 30mm 10mm, clip, width=3.6cm]{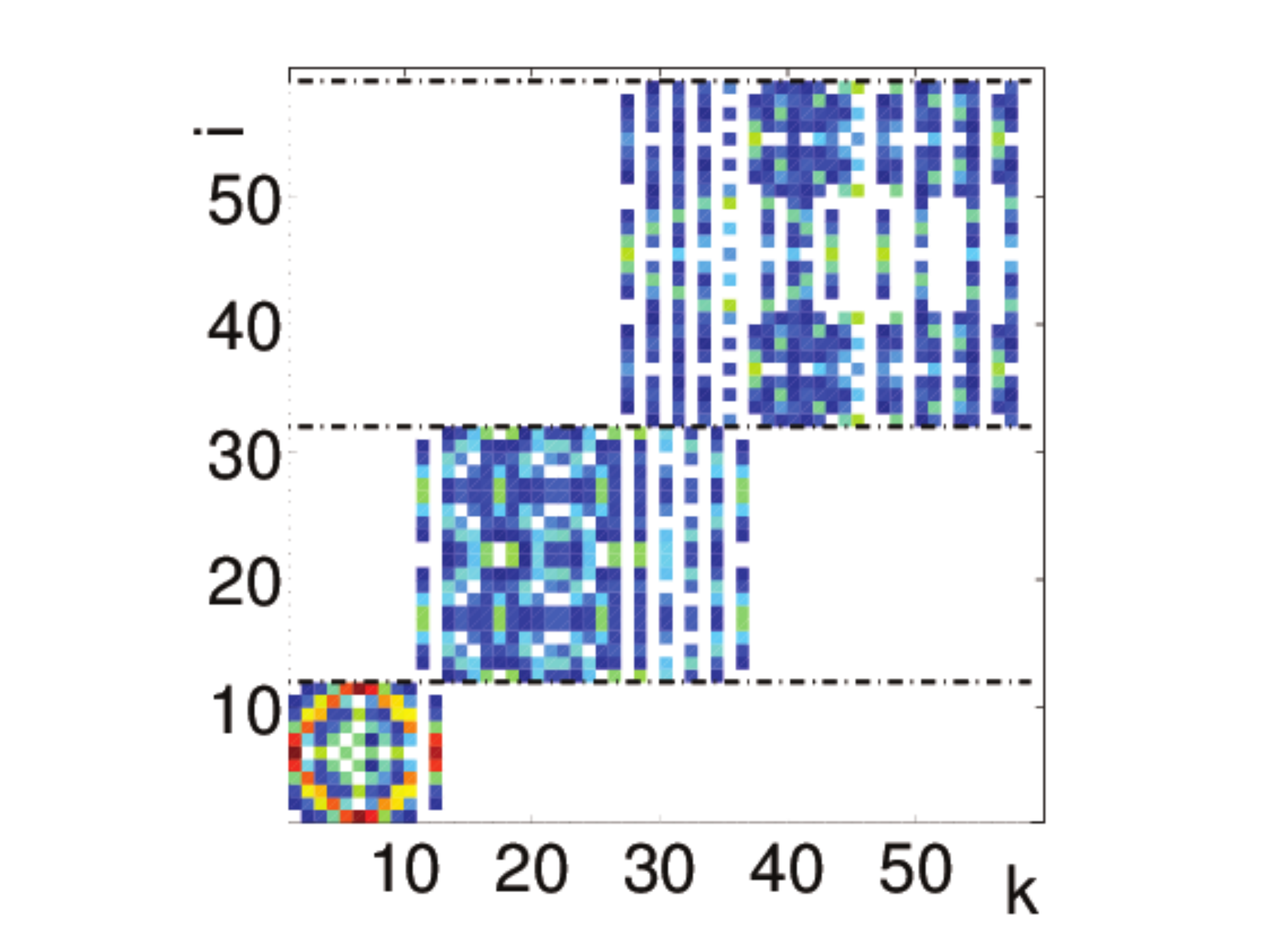} &  \includegraphics[trim =24mm 2mm 30mm 10mm, clip, width=3.8cm]{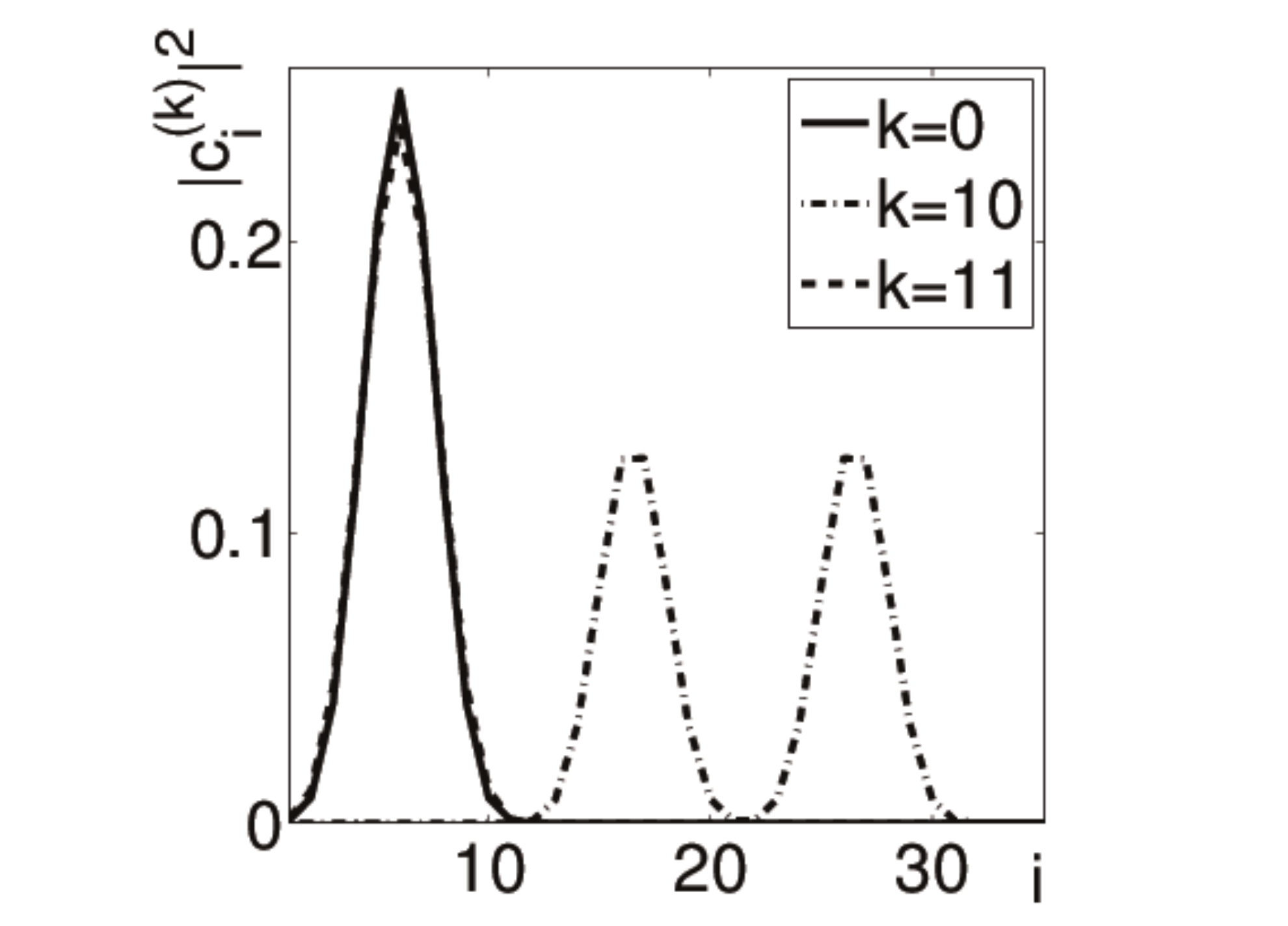}\vspace{-0.2cm} \\
\end{tabular}\\
(d)\\
\includegraphics[trim =35mm 2mm 30mm 10mm, clip,width=6cm]{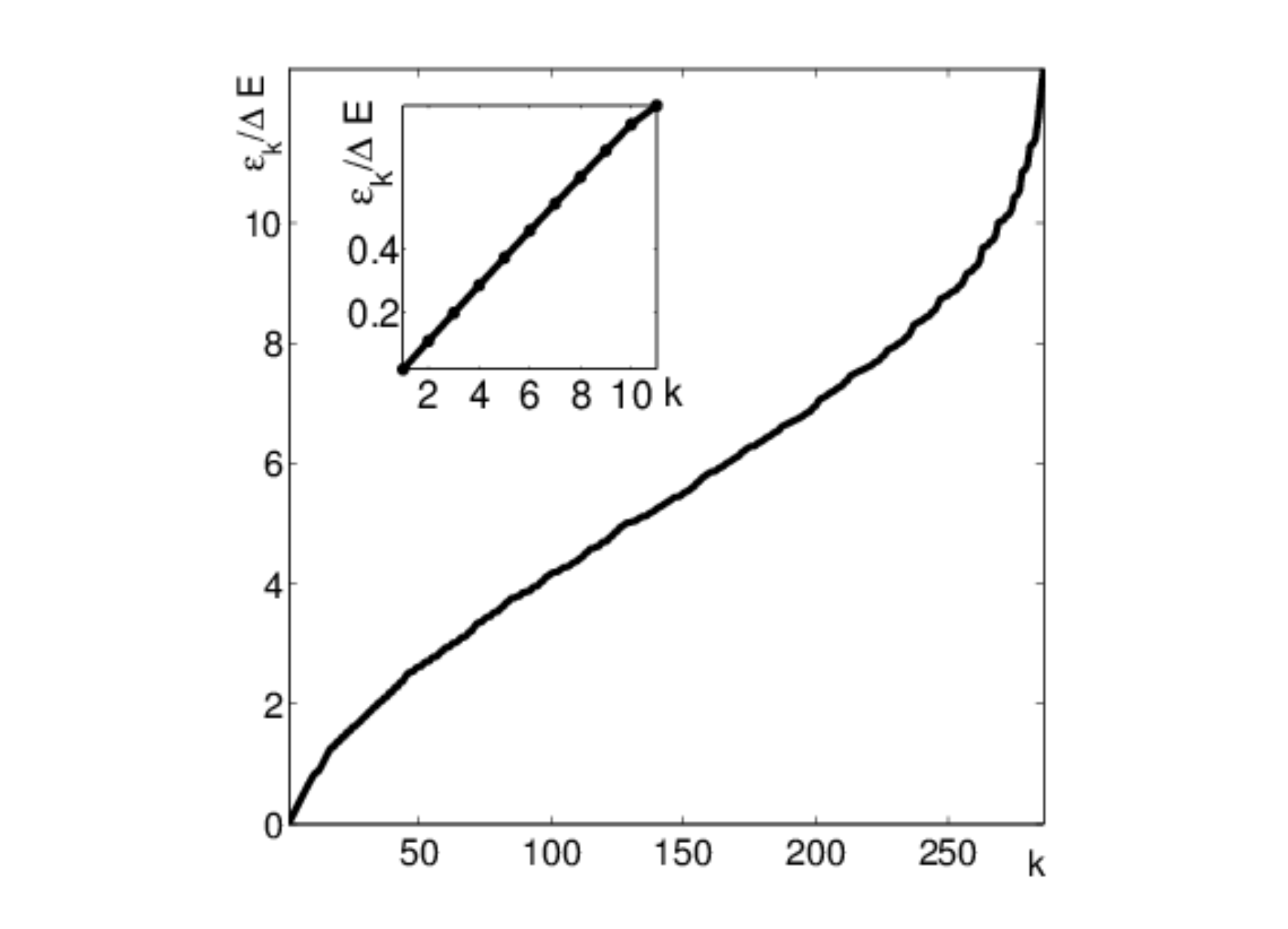}\vspace{-0.3cm}
\end{tabular}
 \vspace{-0.4cm}
 \caption{(Color online)   {\it First crossing in the Josephson regime.} Panel layout as in Fig.~\ref{fig1}. The first crossing in the Josephson regime is induced by
the reduction of $\Delta E$ or  increasing $J_1-J_0$.  \label{fig4}} \end{figure}

\begin{figure}

\begin{tabular}{c}
 \vspace{-0.cm}(a)\\
 \includegraphics[trim =43mm 2mm 30mm 10mm, clip, width=\columnwidth]{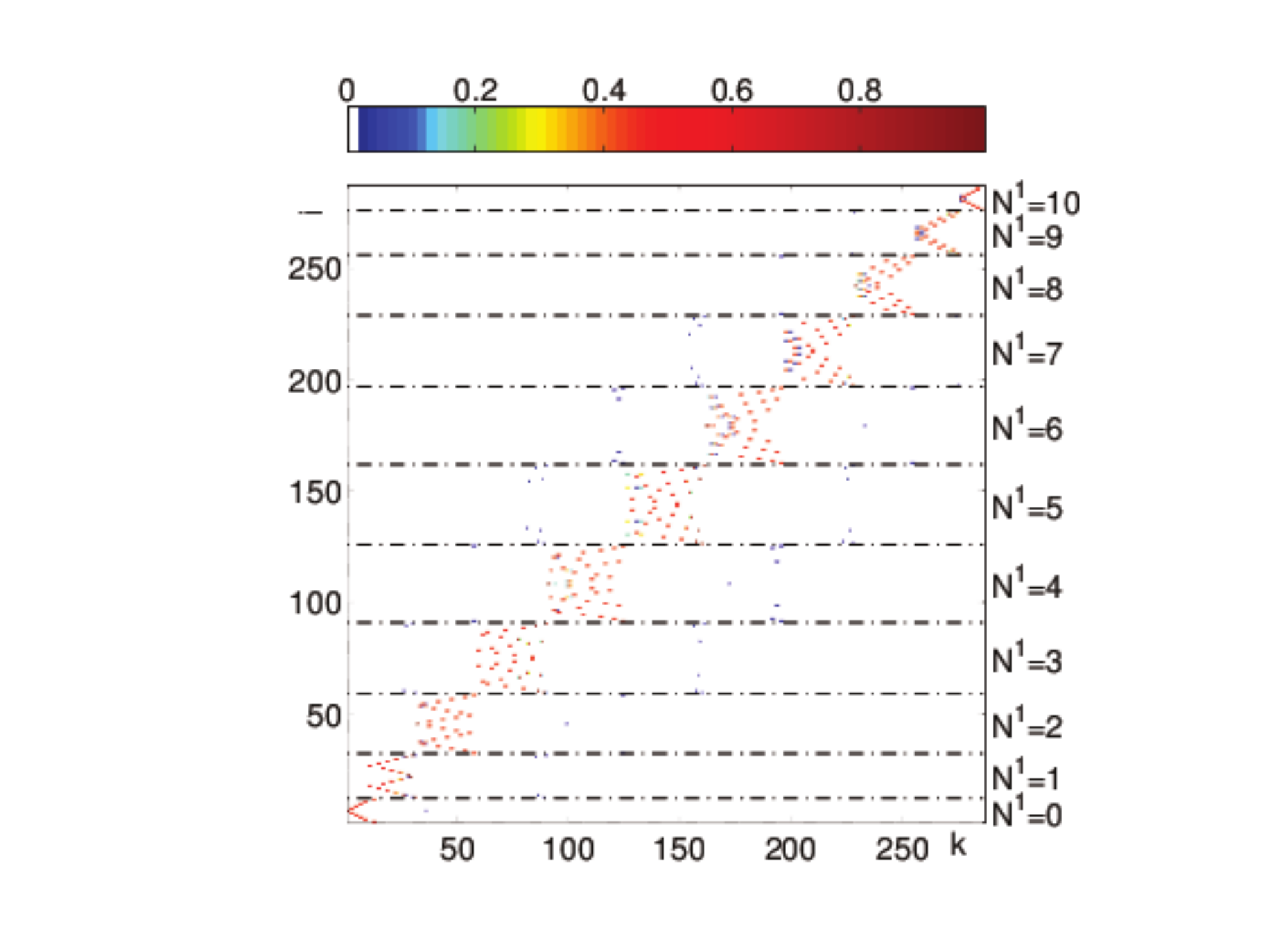}\vspace{-0.6cm} \\
\begin{tabular}{cc}
(b) & (c) \\
\includegraphics[trim =32mm 2mm 30mm 10mm, clip, width=3.6cm]{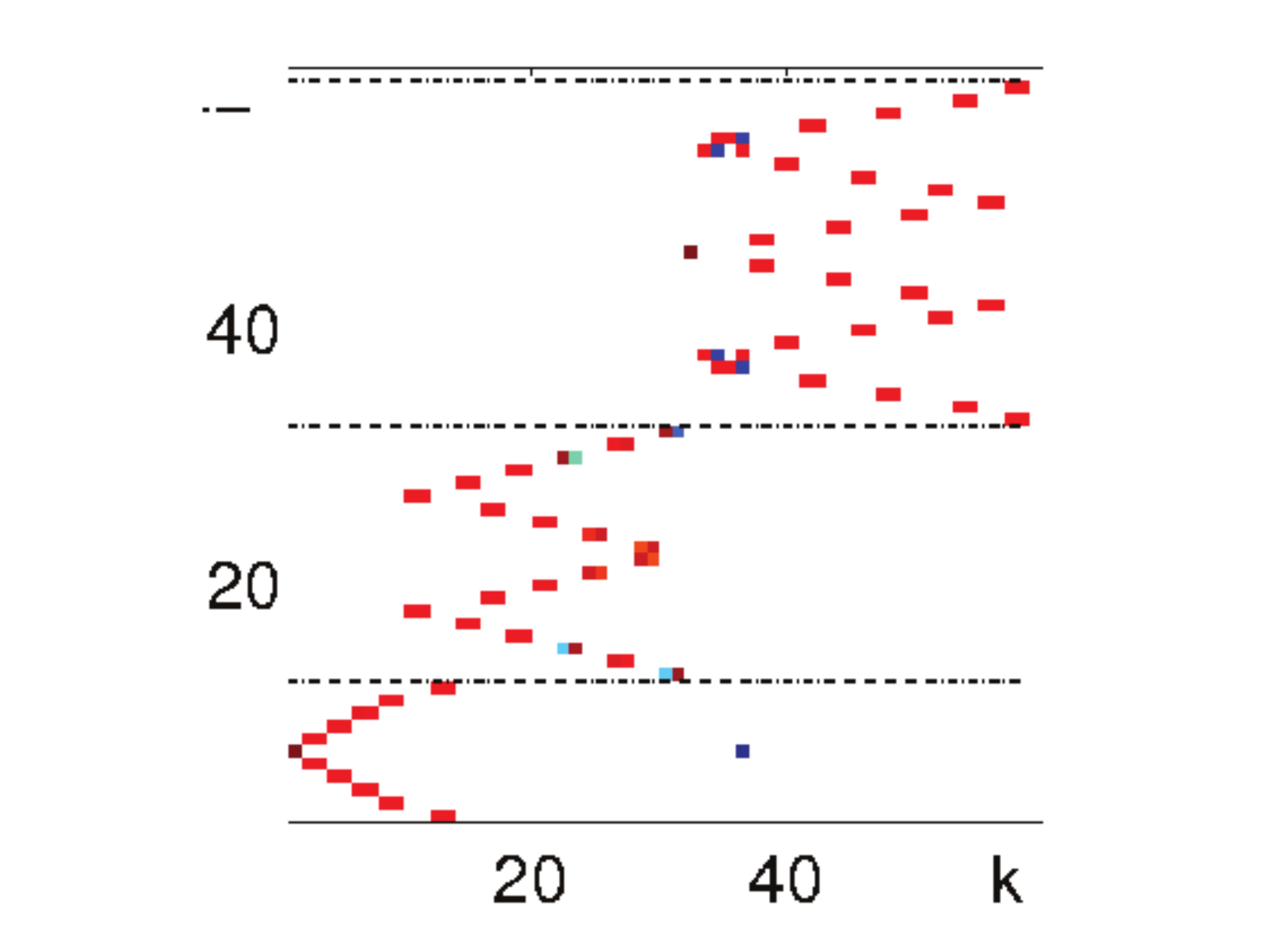} &  \includegraphics[trim =24mm 2mm 30mm 10mm, clip, width=3.8cm]{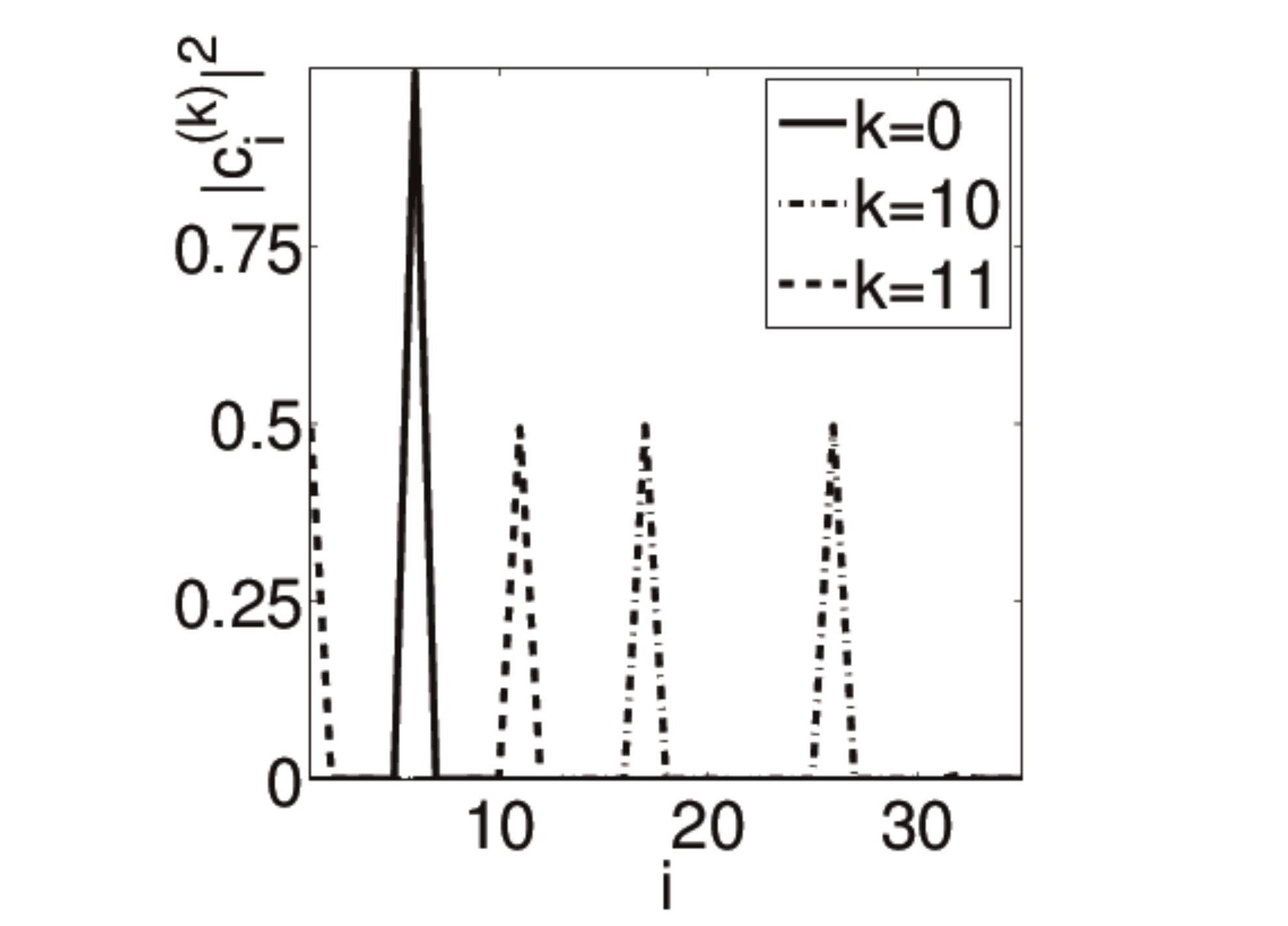} \vspace{-0.2cm}\\
\end{tabular}\\
(d)\\
\includegraphics[trim =35mm 2mm 30mm 10mm, clip,width=6cm]{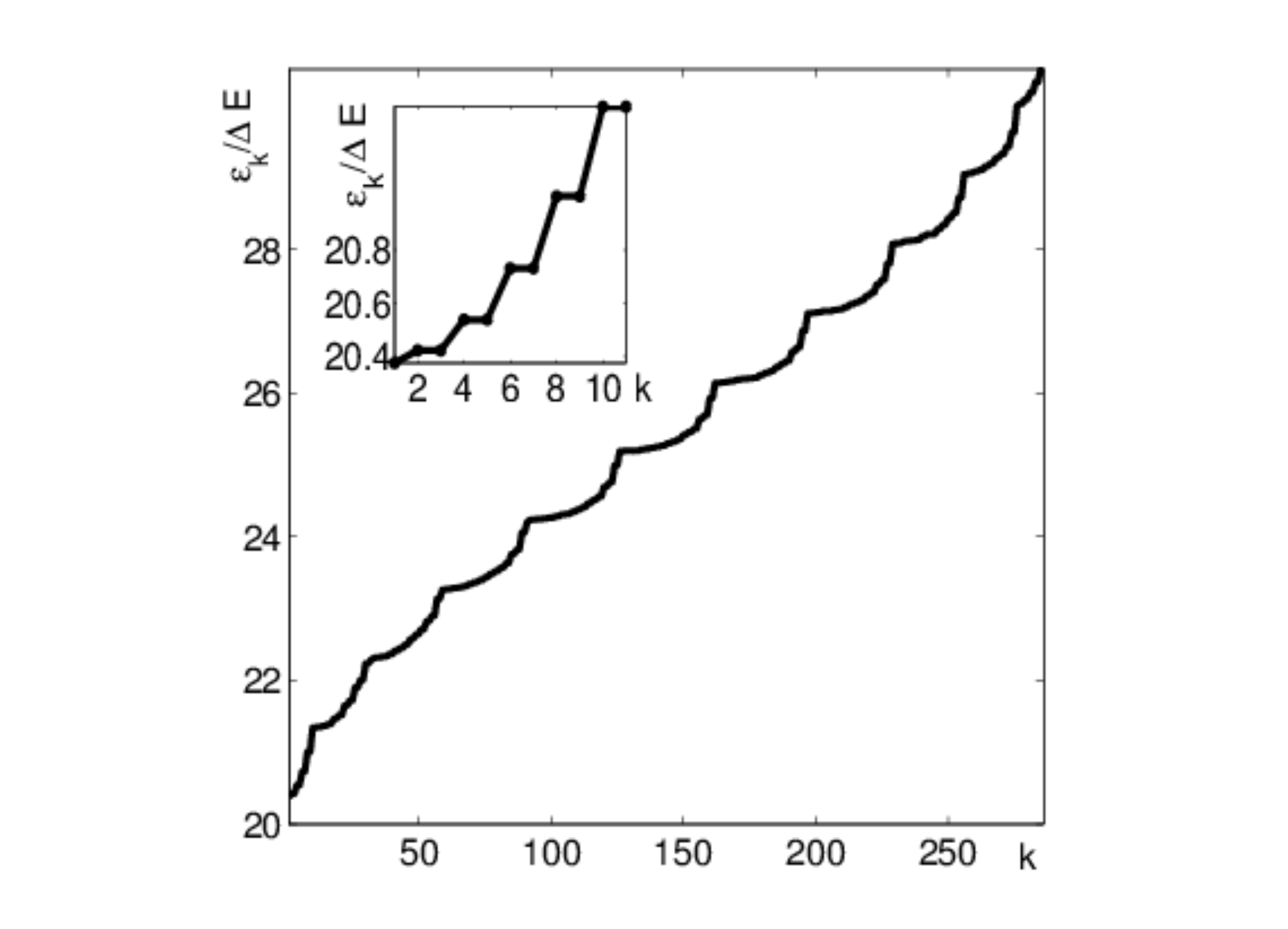}\vspace{-0.3cm}
\end{tabular}
 \vspace{-0.4cm}
  \caption{(Color online)   {\it First crossing in the Fock regime.} Panel layout as in Fig.~\ref{fig1}. The first crossing in this  regime is induced by the reduction
of $\Delta E$ or the increase in $U_0$.  \label{fig5}} \end{figure}

Let us consider  first  the crossings in the Josephson regime. The energy levels are coupled by the interaction energy $U_{01}$. Energy levels only
become completely decoupled when $U_{01}=0$. In this regime $U_{01}$ is small, and then the coupling between levels is weak.
Criterion~(\ref{eq:shad}) gives a good estimate of this coupling. Let us show that, although the energy levels are weakly coupled,
eigenvalue crossings are induced by the presence of the excited level.

We consider first  $U_{0}=0$, $U_{1}=0$ and $U_{01}=0$.  We find that the maximum of the eigenvalues for the first $N+1$
eigenstates with no occupation of the excited level coincides with the minimum of the eigenvalues of the states with one atom in the excited level
if
 \begin{equation}
\chi_{\mathrm{Jos}}=\frac{\triangle E}{J_{0}\,(2N-1)+J_{1}}=1\,.\label{eq:CC}
 \end{equation}
 Equation~(\ref{eq:CC}), determines the first eigenvalue crossing
in this regime. For $\chi_{\mathrm{Jos}}>1$ no crossing occurs. In Fig.~\ref{fig4} we show an example where all the parameters are the same as in Fig.~\ref{fig1},
that is $\zeta_0=10^2$, but with a smaller gap between levels, since $\sigma=235$. As shown in Fig.~\ref{fig4}(b) and (c) the $(N+1)$th eigenvector
shows occupation of the excited level. Notice that for small interactions, the presence of the excited level can induce this crossing.
Fig.~\ref{fig4}(d) shows that  the steps  in the eigenvalues due to the excited atoms observed in Fig.~\ref{fig1}(d)  are now absent, since the
excited atoms have now a much smaller energy. On the other hand, the inset in Fig.~\ref{fig1}(d) shows that the behavior of the eigenvalues is still linear for all the eigenstates which are formed by superpositions of Fock vectors
with the same number of excited atoms.
For
\begin{equation}
\chi_{\mathrm{Jos,gs}}=\frac{\triangle E}{J_{1}-J_{0}}=1\,,\label{eq:CD}
\end{equation}
 the first crossing involving the ground state occurs, i.e., the ground
state is a state with non-zero occupation of the excited level if $\chi_{\mathrm{Jos,gs}}<1$. This is, indeed, a limiting criterion for the validity of the model.

\begin{figure}

\begin{tabular}{c}
 \vspace{-0.cm}(a)\\
 \includegraphics[trim =43mm 2mm 30mm 10mm, clip, width=\columnwidth]{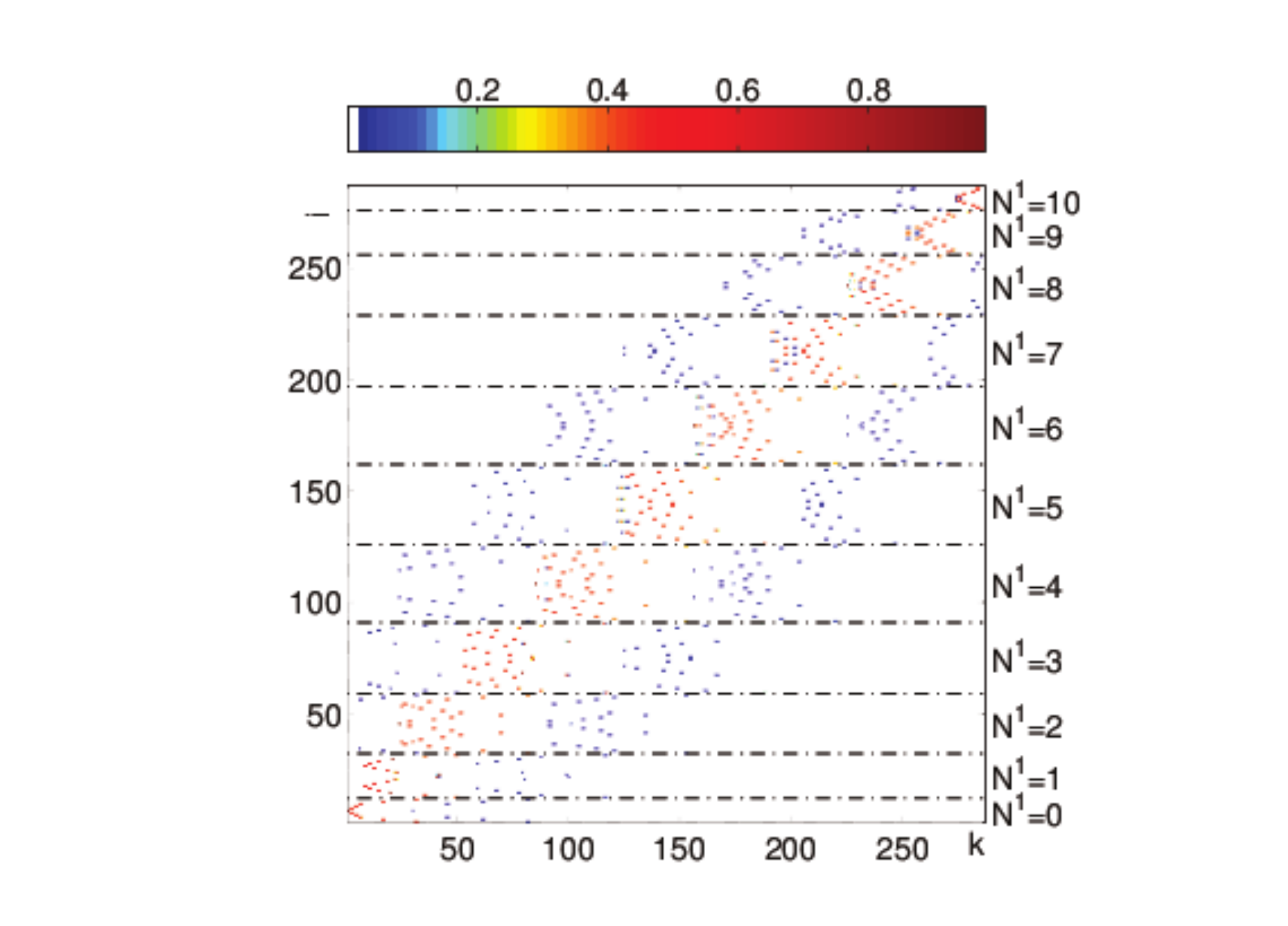}\vspace{-0.6cm} \\
\begin{tabular}{cc}
(b) & (c) \\
\includegraphics[trim =32mm 2mm 30mm 10mm, clip, width=3.6cm]{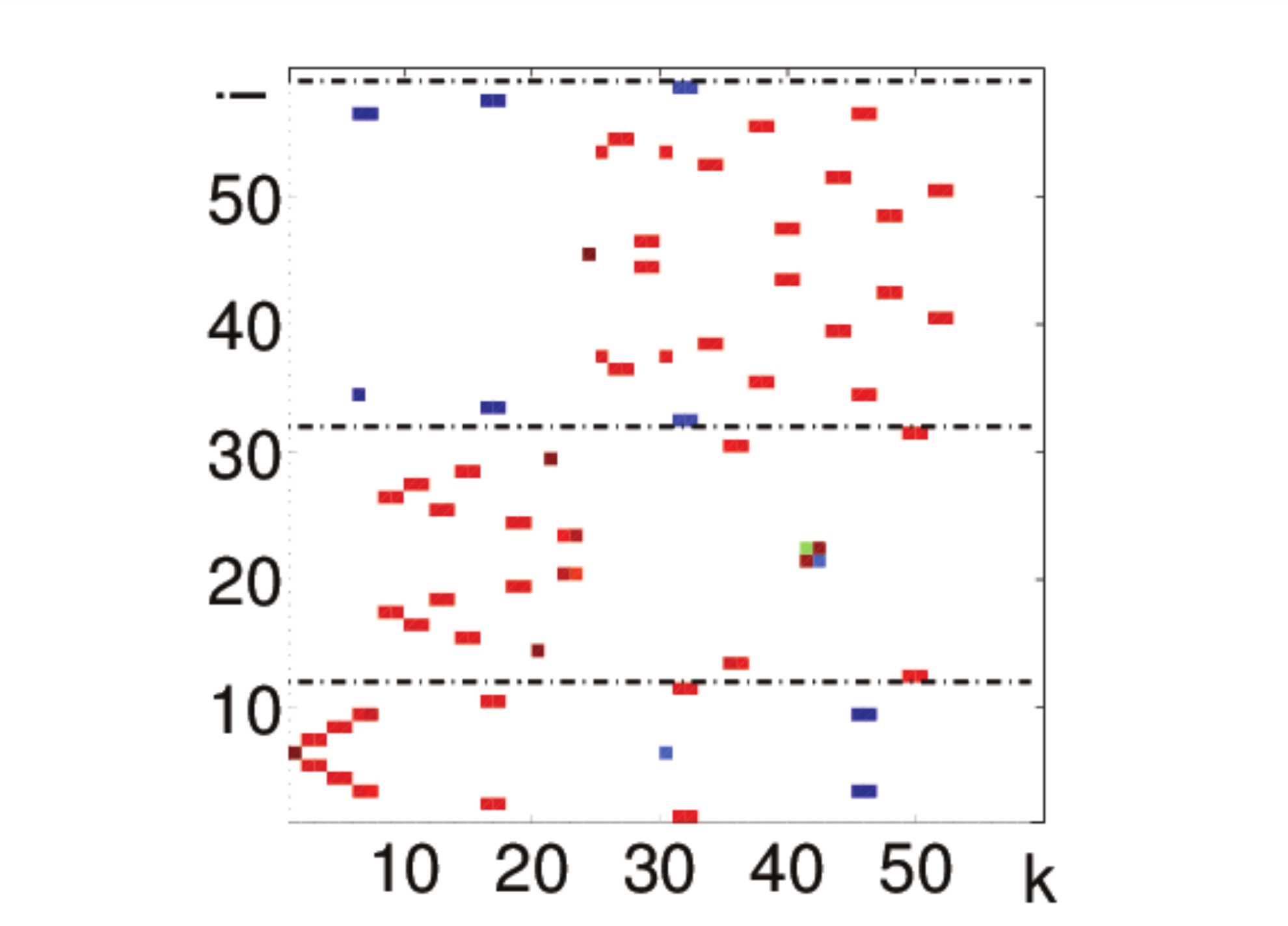} &  \includegraphics[trim =24mm 2mm 30mm 10mm, clip, width=3.8cm]{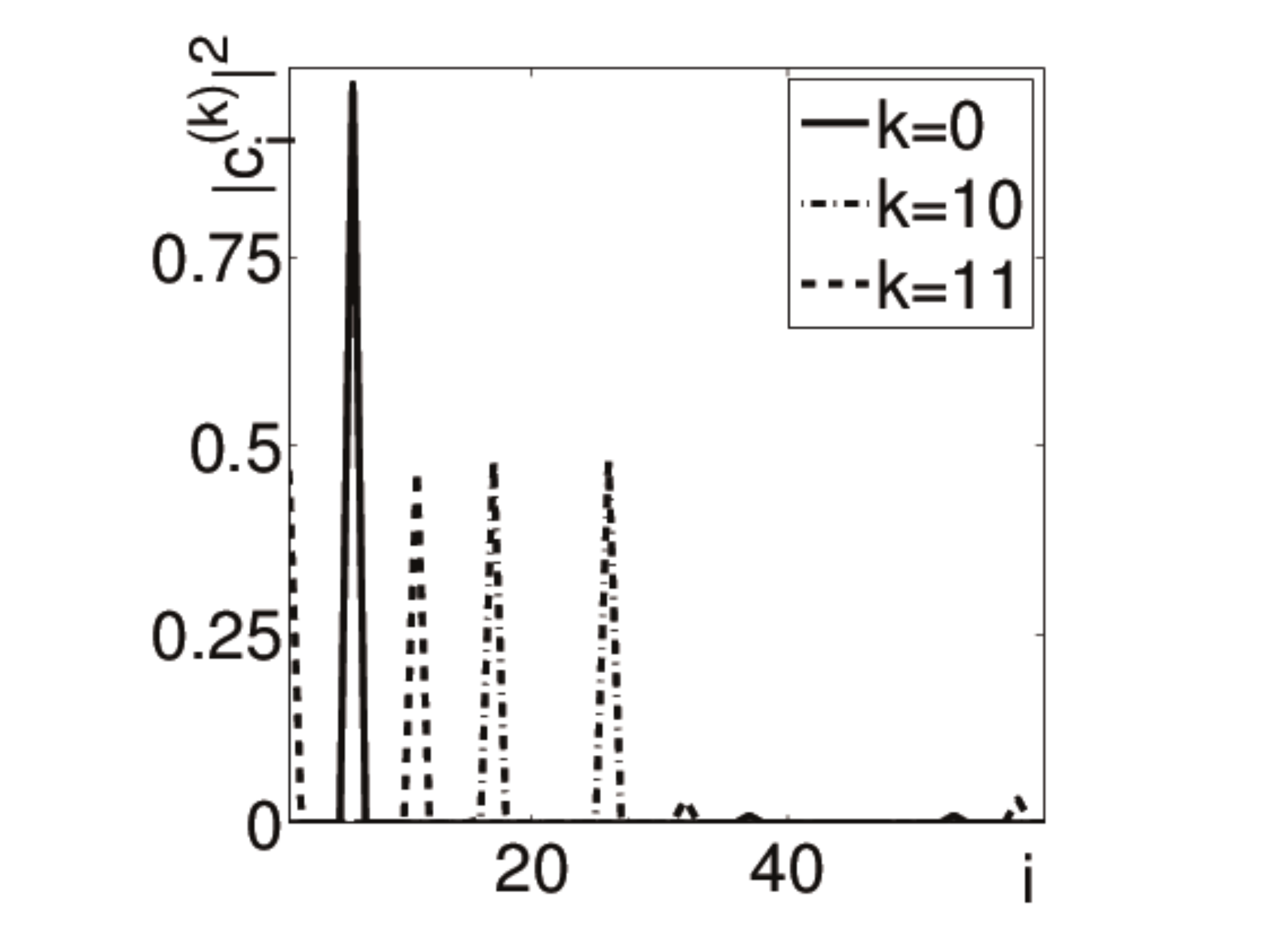}\vspace{-0.2cm} \\
\end{tabular}\\
(d)\\
\includegraphics[trim =35mm 2mm 30mm 10mm, clip,width=6cm]{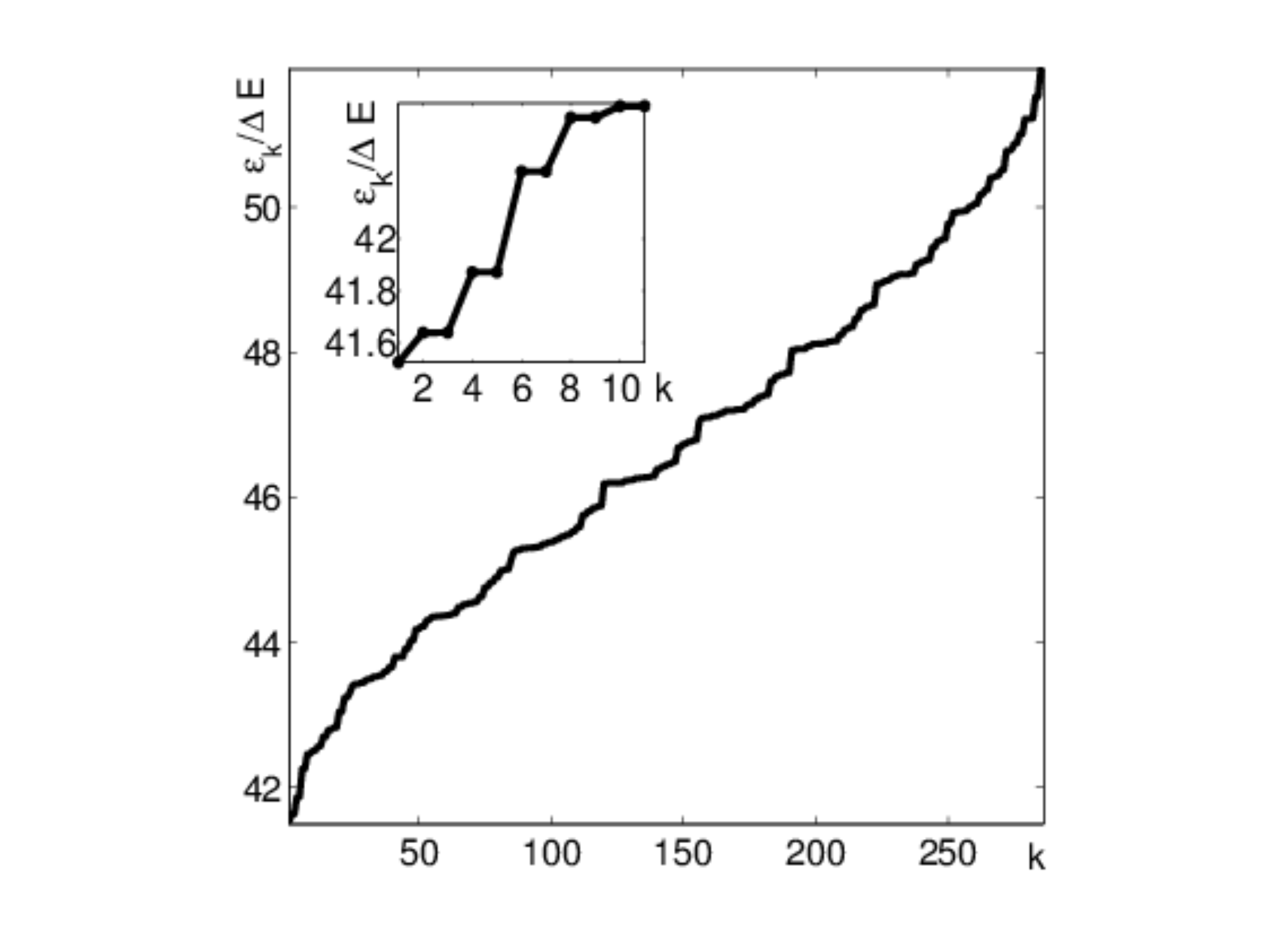}\vspace{-0.3cm}

\end{tabular}
 \vspace{-0.4cm}
 \caption{(Color online)  {\it Shadows of the MS states.} Panel layout as in Fig.~\ref{fig1}. If $U_0$ is
increased, the MS states are coupled to Fock vectors with occupation of the excited level. The eigenvalues still behave as in the Fock regime.
\label{fig6}} \end{figure}

\begin{figure}
\includegraphics[width=\columnwidth]{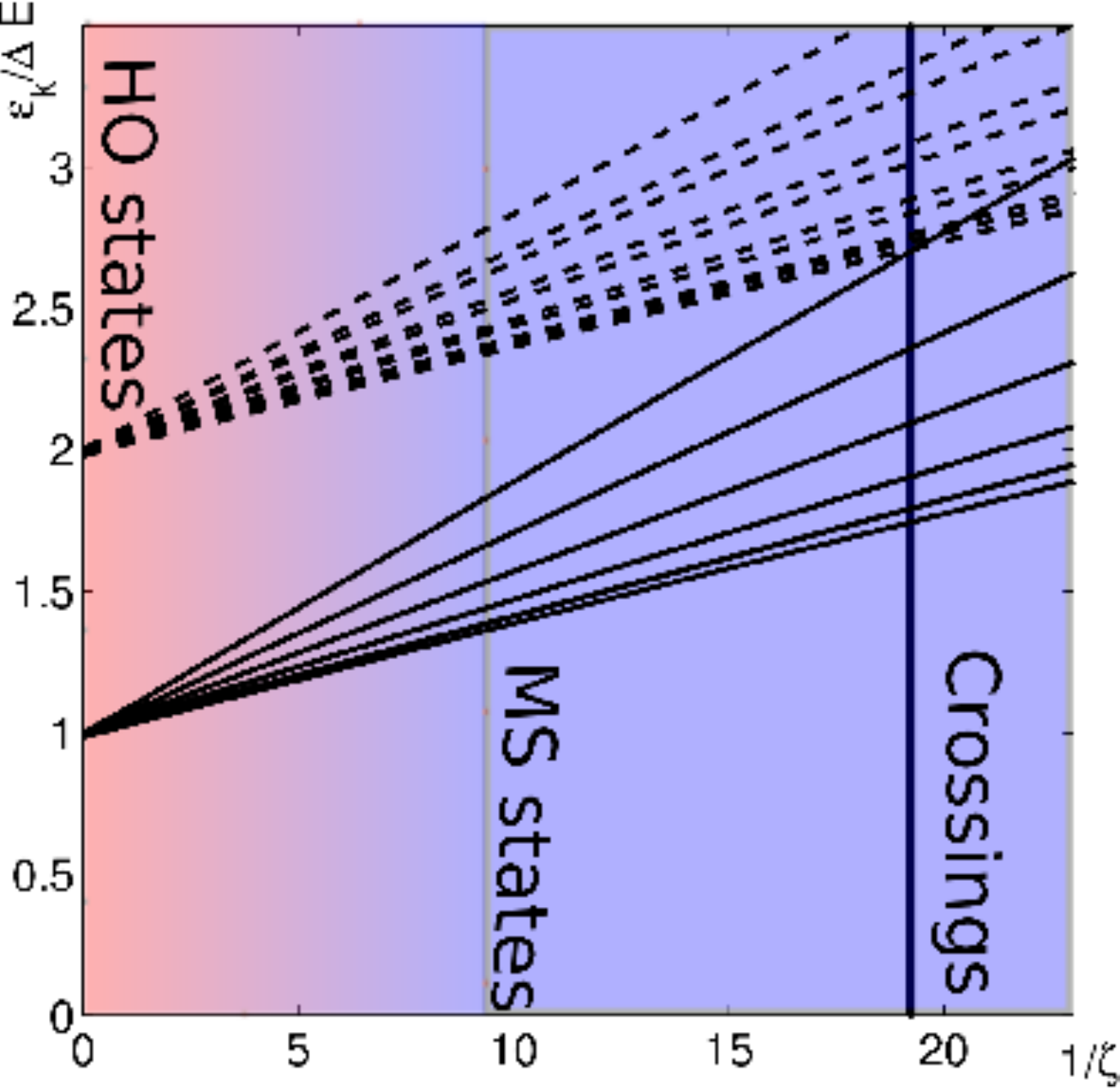}
 \vspace{-0.4cm}
 \caption{(Color online) {\it Crossings as $\zeta$ is
decreased.} Shown are the eigenvalues of the $N+1$ eigenvectors with no occupation of the excited level (solid line) and the  $2(N-1)$ eigenvalues of the eigenvectors showing only one excited atom. The ground state is non-degenerate, while the rest appear in quasidegenerate pairs. The dash-dotted line corresponds to the critical value of $\zeta$ at which
the first crossing occurs, as calculated with~(\ref{eq:CE}). As $\zeta_0$ is decreased, the eigenstates change from the Josephson regime, where they
have are HO states, to the Fock regime, where they are MS states. If $\zeta$ is decreased further, crossings appear, and eventually, shadows  of the MS states. \label{fig7}} \end{figure}

Analogously, in the Fock regime, the first eigenvalue crossings occurs when the condition \begin{equation}
\chi_{\mathrm{Fock}}=\frac{\Delta E}{U_0(N^{2}/2+N-2)-2U_{01}(N/2-1)}\label{eq:CE}
\end{equation}
 is met. This condition is obtained equating the maximum eigenvalue
for states with no occupation of the excited level, to the minimum eigenvalue for states with one particle in the excited level. For $\chi_{\mathrm{Fock}}>1$ no
crossing takes place. In Fig.~\ref{fig5} we show an example for which all the parameters are the same as in Fig.~\ref{fig3}, corresponding to the Fock
regime, but now $\sigma=5.1$. Here, the crossing appeared in the Fock regime, which can be due to the reduction of the gap between levels or to the increasing of the interactions.   Figure~\ref{fig4}(d) and its inset show that the eigenvectors still appear as quasidegenerate
symmetric/antisymmetric MS states, as discussed in Sec.~\ref{sec:MS}.

If we increase further the interactions, or reduce the energy gap, shadows  of the MS states
appear, as discussed at the end of Sec.~\ref{sec:shadows}. For example, if we reduce the $\Delta E$ by taking $\sigma=2.5$, the coupling between the MS states with no excited atoms
and MS states with excited atoms is not negligible, as shown in Fig.~\ref{fig6}(b) and (c).

Finally, the first crossing involving the ground state occurs when the condition
\begin{equation}
\chi_{\mathrm{Fock,gs}}=\frac{\Delta E}{U_0(N-2)-2U_{01}(N/2-1)}\label{eq:CF}
\end{equation}
is satisfied. For $\chi_{\mathrm{Fock,gs}}>1$ the ground state shows non-zero occupation of the
excited level. This criterion should be also considered as a limit for the validity of the model. As discussed in Sec.~\ref{sec:limits}, when this limit is reached  one must consider more energy levels or an MCTDH~\cite{2005MasielloPRA,2006StreltsovPRA,2008AlonPRA} or other many-body method~\cite{2011schollwoeckAP}.

 In Fig.~\ref{fig7}, the first $N+1+2(N-1)$
eigenvalues are represented when $\zeta$ is reduced, for $\sigma=100$. The eigenvectors behave as  HO-like states for certain region, then as MS states, and
finally, the most excited eigenstate with occupation only of the lower level crosses the  least excited state with one atom excited when
criterion~(\ref{eq:CE}) is reached. If $\zeta$ is increased further, shadows  of the MS states appear, and eventually, the limit stated by
criterion~(\ref{eq:CF}) is reached.

Therefore, the bounds to the one level approximation are given by criteria~(\ref{eq:CC}) and (\ref{eq:CE}). We are interested in describing cat-like states which are typically excited eigenstates. Therefore,
a characterization of eigenvalue crossings of energies other than the ground state are also relevant. These crossings appear when criteria~(\ref{eq:CD}) and (\ref{eq:CF}) are satisfied.

\section{Three-dimensional Double Well}
\label{sec:3D}

The three dimensional (3D) double well was extensively studied in~\cite{2011GarciaMarchPRA}. In this case $\omega \sim \omega_{\perp}$. The field operators in Hamiltonian~(\ref{eq:second-quantized}) can be expanded in a fixed well localized single-particle basis obtained from the delocalized eigenfunctions of the single particle Hamiltonian $H_{\mathrm{sp}}=-\frac{\hbar^2}{2\mathcal{M}}\nabla^2 + V(\xvec)$. These eigenfunctions are distortions of the eigenfunctions of the harmonic oscillator potential  $V(\xvec)=\frac{1}{2}\omega^2\xvec^2$, given by
\begin{equation}\label{eq:psi_nlm}
  \psi_{n\ell m}(\xvec) \approx R_{n\ell}(r)Y_{\ell m}(\theta,\phi)\,,
\end{equation}
with $n \in \{0,1,2,\hdots\}$, $\ell \in \{n,n-2,n-4,\hdots,\ell_{\min}\}$, and $m \in \{-\ell,\;-\ell+1,\;\hdots,\;\ell-1,\;\ell\}$.  Here $R_{n\ell}(r)$ is the radial part of the wavefunction, $Y_{\ell m}(\theta,\phi)$ are the familiar spherical harmonics, and $\ell_{\min}$ is 0 when $n$ is even and 1 when $n$ is odd.  Then,  $n$ is the single-particle energy level, $\ell$ is the orbital angular momentum in 3D, and $m$ is its $z$-projection. For the two level approximation  $n=\ell$, and therefore the index $n$ is superfluous and hereafter suppressed.   These functions can be explicitly written in spherical coordinates for $\ell=0,1$ as
\begin{align*}
\psi_{00}(r,\theta,\varphi)&=a_{\mathrm{ho}}^{-3/2}\pi^{-3/4}e^{-r^{2}/2a_{\mathrm{ho}}^2}\,,\\
\psi_{10}(r,\theta,\varphi)&=a_{\mathrm{ho}}^{-3/2}\sqrt{2}\pi^{-3/4}(r/a_{\mathrm{ho}})e^{-r^{2}/2a_{\mathrm{ho}}^2}\cos(\theta)\,,\\
\psi_{1\pm1}(r,\theta,\varphi)&=a_{\mathrm{ho}}^{-3/2}\pi^{-3/4}(r/a_{\mathrm{ho}})e^{-r^{2}/2a_{\mathrm{ho}}^2}\sin(\theta)e^{\pm i\varphi}\,. \end{align*}
The energy of an atom associated with the wavefunction $\psi_{\ell m}(\xvec-\xvec_j)$ is
\begin{equation}\label{eq:E_nlm3}
  E_{n} \approx \hbar\omega(n+3/2)\,.
\end{equation}
In general, the spherical harmonics cannot be accurately used as a basis.  In particular, they give poor approximations of the  overlap between functions localized in different wells. Then, the localized functions at well $j$ can be obtained numerically  as
\begin{align}
\psi_{j00}(\xvec)&=\phi(x)^0\phi(y)^0\psi_{j0}(z)\,,\\
\psi_{j10}(\xvec)&=\phi(x)^0\phi(y)^0\psi_{j1}(z)\,,\\
\psi_{j,1\pm1}(r,\theta,\varphi)(\xvec)&= \frac{1}{\sqrt{2}}\left[\phi^{1}(x)\phi^{0}(y)\psi_{j0}(z)\right.\\
&\pm \left. i\phi^{0}(x)\phi^{1}(y)\psi_{j0}(z)\right]\,, \end{align}
where $\phi^{\ell}(x)$ and $\phi^{\ell}(y)$ are two lowest excited eigenfunctions of the harmonic oscillator in the $x$ and $y$ directions, and  $\psi_{j\ell}(z)$ are the on-well localized eigenfunctions of the double well given in~(\ref{eq:sup1}) and~(\ref{eq:sup2}), respectively. The four modes localized in one of the wells are represented schematically in Fig.~\ref{fig0}(b).

Then, we can expand the field operator in~(\ref{eq:second-quantized1}) in terms of this eight mode basis as
\begin{equation}\label{eq:hatPsi3}
  \hat{\Psi}(\xvec)
  = \sum_{j,\ell,m}\bhatp{j}{\ell}{m}{}\psi_{\ell m}(\xvec-\xvec_j)\,,
\end{equation}
where  $\xvec_{1} \equiv -\xvec_{\mathrm{min}}$ and $\xvec_2 \equiv \xvec_{\mathrm{min}}$ are the minima of the left and right wells. The operators $\bhatp{j}{\ell}{m}{\dagger}$ and $\bhatp{j}{\ell}{m}{}$ satisfy the usual bosonic annihilation and creation commutation relations,
\begin{align}
[\bhatp{j}{\ell}{m}{},\;\bhatp{j'}{\ell'}{m'}{\dagger}]&=\delta_{jj'}\delta_{\ell\ell'}\delta_{mm'}\,,\nonumber\\
[\bhatp{j}{\ell}{m}{\dagger},\;\bhatp{j'}{\ell'}{m'}{\dagger}]&=[\bhatp{j}{\ell}{m}{},\;\bhatp{j'}{\ell'}{m'}{}]=0\,.
\end{align}

By using this eight-mode expansion of the field operator,  the 3D Hamiltonian was obtained:
\begin{equation}
\hat{H} = \sum_{\ell,m} \hat{H}_{\ell m} + \hat{H}_{\mathrm{int}}\, .
\end{equation}
 The first term is a sum over LMGHs for each level, similar to the one discussed in Sec.~\ref{sec:1DLMG} [See Eq.~(\ref{Eq:onelevelH})]:
 \begin{align}
\label{eq:LMGHpart}
H_{\ell m}&= U_{\ell m}\sum_j  \nhatp{j}{\ell}{m}(\nhatp{j}{\ell}{m}-1)-J_{\ell m}\sum_{j'\neq j} \bhatp{j}{\ell}{m}{\dagger}\bhatp{j'}{\ell}{m}{}\nonumber\\
&+E_{\ell}\sum_j\nhatp{j}{\ell}{m}\,.
 \end{align}
 In Eq.~(\ref{eq:LMGHpart}), the term  $E_{\ell}\sum_j\nhatp{j}{\ell}{m}$ accounts for the energy of the atoms at level $\ell$  and $z$ component of angular momentum $m$, with $\nhatp{j}{\ell}{m}=\bhatp{j}{\ell}{m}{\dagger}\bhatp{j}{\ell}{m}{}$ the number operator. The term
\begin{equation}
\label{eq:Hinter}
\hat{H}_{\mathrm{int}} =\sum_m \hat{H}_{\mathrm{inter}}^m+\hat{H}_{\mathrm{intra}}\,,
\end{equation}
accounts for processes among atoms in different  levels and with different $z$ component of the angular momentum.

 There are three relevant processes in Eq.~(\ref{eq:Hinter}). The first of these  is given by
\begin{align}
\label{eq:Hinter1}
\hat{H}_{\mathrm{inter}}^0  & = \!\sum_{j}  \Big{\{}U_{0 1}^{0 0}    \left[\!\left(\!\bhatp{j}{0}{0}{\dagger}\!\right)^2\!
     \left(\bhatp{j}{1}{0}{}\right)^2+\hc\right]\nonumber\\
&+4\,U_{0 1}^{0 0}\,\nhatp{j}{0}{0}\,\nhatp{j}{1}{0}\Big{\}}\,.
\end{align}
This process describes the excitation of two atoms from  the ground state to an orbital with $m=0$, or  conversely, their decay from an excited state with $m=0$ to the ground state.  We name this process {\it zero-vorticity interlevel hopping}. The second is
\begin{align}
\label{eq:Hinter2}
\hat{H}_{\mathrm{inter}}^1 & = \!\sum_{j} \Big{\{} U_{0 1}^{0 1}    \left[\!\left(\bhatp{j}{0}{0}{\dagger}\!\right)^2
     \bhatp{j}{1}{1}{}\bhatp{j}{1,}{-1}{}+\hc\right]\nonumber\\
& +4\,U_{0 1}^{0 1}\,(\nhatp{j}{0}{0}\,(\nhatp{j}{1}{1}+\nhatp{j}{1,}{-1}))\Big{\}}\,.
\end{align}
This process is  similar to the first one and we call it   {\it vortex-antivortex interlevel hopping}. It  permits an atom to change its level and also its $z$ component of its angular momentum, $m$. Through this process two atoms in the ground  state can be  excited, one with $m=1$ (a vortex) and the other with $m=-1$ (an anti-vortex). Conversely, one atom with $m=+1$ and another with $m=-1$ can decay to the ground state. Finally, the third process is
\begin{align}
\label{eq:Hinter3}
\hat{H}_{\mathrm{intra}} & = \!\sum_{j} \Big{\{}U_{1 1}^{0 1}\
  \left[\!\left(\!\bhatp{j}{1}{0}{\dagger}\!\right)^2\!
  \bhatp{j}{1}{1}{}\bhatp{j}{1}{,-1}{}
  +\hc\right]\\
& + 2\,U_{1 1}^{0 1}\,(\nhatp{j}{1}{0}\,(\nhatp{j}{1}{1}+\nhatp{j}{1,}{-1})
+2\,U_{11}\,(\nhatp{j}{1}{1}\nhatp{j}{1,}{-1})\Big{\}}\,.\nonumber
\end{align}
Through the third process,  {\it vortex-antivortex intralevel hopping},  the atoms can only change their angular properties, but not their energy level. Then, two  excited atoms with $m=0$ can generate a pair of atoms, one with $m=1$ and the other with $m=-1$ or vice versa.  These three processes are represented schematically in Fig.~\ref{fig0}(b).  Again, in obtaining this Hamiltonian, the off-site interaction coefficients were neglected, as discussed in  Sec.~\ref{sec:1DLMG} for the 1D case.

Notice that the 3D interaction and tunneling coefficients that appear in the LMGH part of this Hamiltonian, Eq.~(\ref{eq:LMGHpart}), have to be evaluated independently for atoms in the same level and with the same $m$. The tunneling coefficients in Eq.~(\ref{eq:LMGHpart}) are defined as
\begin{align}
 J_{\ell m}&=-\int d^3\xvec\psi^{\ast}_{\ell m}(\xvec-\xvec_{\mathrm{min}}) \left[-\frac{\hbar^{2}}{2\mathcal{M}}\nabla^{2}+V(\xvec)\right] \nonumber\\
           &\times \psi_{\ell m}(\xvec+\xvec_{\mathrm{min}})\,.\label{eq:J_lm}
\end{align}
The interaction coefficients are
\begin{equation}
U_{\ell m}^{\ell' m'}=\frac{g}{2}\int d^3\xvec|\psi_{\ell m}(\xvec)|^{2}|\psi_{\ell' m'}(\xvec)|^{2}\,.
\label{eq:U_lm_l'm'}
\end{equation}
Notice that the coefficients in the coupling part of the Hamiltonian,  Eqs.~(\ref{eq:Hinter1})-(\ref{eq:Hinter3}), are related to the  interaction coefficients, and not to tunneling coefficients.

The presence of many new coefficients makes necessary a wider characterization of regimes, further away from the 1D Josephson, intermediate, and Fock regimes (where the eigenvectors and eigenvalues behave as in  Figs.~\ref{fig1}, \ref{fig2}, and \ref{fig3}).  This was accomplished in~\cite{2011GarciaMarchPRA}, where criteria for the crossings and limits of the model were also given, and represented numerically for the particular case of the Duffing double well potential $V(z)=V_0(-8 z^2/a^2+16 z^4/a^4+1)$.   A Josephson and Fock regime can be described, for which the eigenvectors can be characterized as HO-like or MS states, respectively.  These MS states can show occupation of the excited level, resulting in an MS of angular degrees of freedom. The presence of these orbitals are the fundamental difference from the 1D case. Moreover, the three coupling processes described above give rise to interesting dynamical phenomena, like vortex tunneling and vortex/antivortex creation/annihilation along  with the tunneling of atoms with non-zero  $m$.  These new dynamical phenomena are the subject of  our upcoming work~\cite{2011GarciaMarchpre}.

\section{Conclusions}
\label{sec:conclusion}

Ultracold bosons in double well potentials are a simple system to study a great variety of physical phenomena. There are two main  processes: the atoms can interact in pairs on-well with energy $U_0$  or tunnel to the other well with energy $J_0$.  When the interactions dominate over the tunneling energies in the system, the  spectra of the eigenvectors are characterized by the presence of MS states. But also in this case it is necessary to consider the possibility that the atoms populate an excited level.  Then, other energies are relevant,  including the interaction energy of the atoms in the excited level $U_1$, the tunneling energy in the excited level $J_1$, and the  single-particle energy gap between levels, $\Delta E$.

 In this Article, we used a two level approach to describe the possible physical scenarios in one- and three-dimensional double wells. For the 1D case, we clearly  identified the Josephson regime, for which quantum tunneling was experimentally demonstrated.  We characterized the eigenvectors and eigenvalues in the noninteracting regime, and we showed through direct diagonalization of the two level LMGH that this behavior can be extended to the Josephson regime. Our interest was mainly in the occurrence of MS states of atoms localized in either one or the other well, which appear for bigger interactions, in the Fock regime.    We characterized the appearance of crossings in the spectra, which serves as a limiting criterion for the validity of the  one-level or LMGH approximation. We obtained also the limits of the  two-level approximation. In particular, we found when the interactions are sufficiently large that coupling to states with non-zero occupation of the excited level is relevant. In this case, the  interaction coefficient can overcome the energy gap between levels, and then the states show non-zero occupation of the excited level.  We described numerically the transition from the Josephson regime to the  Fock regime. We introduced a new Josephson-Fock \emph{mixed regime},  in which the most excited eigenvectors with no occupation of the excited level behave as MS states while the less excited ones are HO-like states.  Moreover, since $J_1\gg J_0$, when the eigenvectors involve only atoms in the excited level, they behave as  HO-like states.

MS states are, unfortunately, highly excited states and fragile against decoherence processes~\cite{2000DalvitPRA,2010PichlerPRA}.   Thus they are difficult to observe experimentally. Quantum superpositions of matter waves have been observed for few particles, like electrons, but remains a challenging problem for larger objects; experiments with $C_{60}$ molecules are in the lead at present in such efforts (see~\cite{1999ArndtNature,2002BrezgerPRL} and references therein), but ultracold bosons have the potential to go to hundreds or thousands of particles in an MS state.  Therefore, there are many theoretical proposals  for realizing  them in  a BEC experiment~\cite{2001MenottiPRA,2004HigbiePRA,2005MahmudPRA,2006HuangPRA,2006DunninhamNJP,2008PiazzaPRA,2008FerriniPRA,2008MazetsEPL,2010watanabePRA,2010XuPRA,2010HallwoodPRA,2010CarrEPL,2011KanamotoJO},
 and this  remains an appealing research topic with deep physical implications.   The two-level scenario introduces the possibility to study other initial states with non-zero occupation of the excited level, which could be the key to realize MS states experimentally, as discussed in~\cite{2010CarrEPL}.  Also, the  two-level approach in three dimensions permits one to consider angular degrees of freedom in the problem~\cite{2011GarciaMarchPRA}.  The angular degrees of freedom are a very interesting topic in systems of ultracold atoms in optical lattices~\cite{2005IsacssonPRA,2005ScarolaPRL,2006LiuPRA,2007XuPRB,2010CollinPRA,2011LewensteinNatPhys}, where the excitation of part of the population to a $p$ level was demonstrated experimentally~\cite{2007MullerPRL,2011WirthNatPhys}. Here, we summarized the derivation of the Hamiltonian and the MS states. Then, the study of the dynamics of ultracold atoms  in three dimensional double wells where  orbital degrees of freedom play a relevant role is an interesting topic for future research.

\acknowledgements{ This material is based in part upon work supported by the National Science Foundation under  grant numbers PHY-0547845 and PHY-1067973. L.D.C. thanks the Alexander von Humboldt foundation  and the Center for Quantum Dynamics for additional support. M.A.G.M. acknowledges  the Fulbright Foundation, FECYT, and the  Spanish Ministry of Education and Science.  }

\end{document}